\RequirePackage{fix-cm}
\documentclass[twocolumn]{svjour3}

\smartqed  

\usepackage{amsmath,amssymb,latexsym,amsfonts,mathrsfs}
\usepackage{graphicx}
\usepackage{pdfpages}
\usepackage{float}
\usepackage[ruled]{algorithm2e}
\usepackage{xparse} 
\usepackage[compatibility=false]{caption}
\usepackage{subcaption}
\usepackage[margin=1in]{geometry}
\usepackage{verbatim}
\usepackage{mathtools}
\usepackage[hyphens]{url}

\usepackage{lscape}
\DeclareMathOperator*{\argmax}{arg\,max}

\ExplSyntaxOn
\NewDocumentCommand{\stringprocess}{ m m }
 {
  \egreg_string_process:nn { #1 } { #2 }
 }
\cs_new_protected:Npn \egreg_string_process:nn #1 #2
 {
  \tl_map_inline:nn { #2 } { #1 { ##1 } }
 }
\ExplSyntaxOff
\newcommand{\negspace}[1]{#1\hspace{-0.06cm}} 
\newcommand{\mat}[1]{\stringprocess{\negspace}{#1}\hspace{0.08cm}}

\newcommand{\vz}[1]{\ensuremath{\mathbb{#1}}}

\newcommand{\R}{{\vz R}}

\begin{document}

\title{Unsupervised record matching with noisy and incomplete data
}

\author{Yves van Gennip        \and
        Blake Hunter \and
				Anna Ma\and
				Daniel Moyer\and
				Ryan de Vera\and
				Andrea L. Bertozzi
}

\institute{Y. van Gennip \at
              University of Nottingham \\
              \email{Y.VanGennip@nottingham.ac.uk}           
           \and
           B. Hunter \at
              Claremont McKenna College\\
							\email{bhunter@cmc.edu}
					\and
							A. Ma \at
							Claremont Graduate University\\
							\email{anna.ma@cgu.edu} 
					\and
							D. Moyer \at
							University of Southern California\\
							\email{moyerd@usc.edu} 
					\and
							R. de Vera \at
							formerly California State University, Long Beach\\
							\email{ryan.devera.03@gmail.com}
					\and
							A. L. Bertozzi \at
							University of California, Los Angeles\\
							\email{bertozzi@math.ucla.edu}
}

\date{}

\maketitle

\begin{abstract}
We consider the problem of duplicate detection in noisy and incomplete data: given a large data set in which each record has multiple entries (attributes), detect which distinct records refer to the same real world entity. This task is complicated by noise (such as misspellings) and missing data, which can lead to records being different, despite referring to the same entity. Our method consists of three main steps: creating a similarity score between records, grouping records together into ``unique entities'', and refining the groups. We compare various methods for creating similarity scores between noisy records, considering different combinations of string matching, term frequency-inverse document frequency methods, and $n$-gram techniques. In particular, we introduce a vectorized soft term frequency-inverse document frequency method, with an optional refinement step. We also discuss two methods to deal with missing data in computing similarity scores.

We test our method on the Los Angeles Police Department Field Interview Card data set, the Cora Citation Matching data set, and two sets of restaurant review data. The results show that the methods that use words as the basic units are preferable to those that use $3$-grams. Moreover, in some (but certainly not all) parameter ranges soft term frequency-inverse document frequency methods can outperform the standard term frequency-inverse document frequency method. The results also confirm that our method for automatically determining the number of groups typically works well in many cases and allows for accurate results in the absence of a priori knowledge of the number of unique entities in the data set.

\keywords{ duplicate detection \and data cleaning \and data integration \and record linkage \and entity  matching \and identity uncertainty \and transcription error}

\end{abstract}

\section{Introduction}
\label{intro}
Fast methods for matching records in databases that are similar or identical have growing importance as database sizes increase \cite{winkler1999state,Winkler06overviewof,elmagarmid2007duplicate,manning2008introduction,ahmed2010dynamic}. Slight errors in observation, processing, or entering data may cause multiple unlinked nearly duplicated records to be created for a single real world entity. Furthermore, records are often made up of multiple attributes, or fields; a small error or missing entry for any one of these fields could cause duplication.

For example, one of the data sets we consider in this paper is a database of personal information generated by the Los Angeles Police Department (LAPD). Each record contains information such as first name, last name, and address. Misspellings, different ways of writing names, and even address changes over time, can all lead to duplicate entries in the database for the same person.

Duplicate detection problems do not scale well. The number of comparisons which are required grows quadratically with the number of records, and the number of possible subsets grows exponentially. Unlinked duplicate records bloat the storage size of the database and make compression into other formats difficult. Duplicates also make analyses of the data much more complicated, much less accurate, and may render many forms of analyses impossible, as the data is no longer a true representation of the real world. After a detailed description of the problem in Section~\ref{sec:terminology} and a review of related methods in Section~\ref{sec:existingmethods}, we present in Section~\ref{sec:newmethod} a vectorized \emph{soft term frequency-inverse document frequency} (soft TF-IDF) solution for string and record comparison. In addition to creating a vectorized version of the soft TF-IDF scheme we also present an automated thresholding and refinement method, which uses the computed soft TF-IDF similarity scores to cluster together likely duplicates. In Section~\ref{sec:Applications} we explore the performances of different variations of our method on four text databases that contain duplicates.

\section{Terminology and problem statement}\label{sec:terminology}

We define a data set $D$ to be an $n \times a$ array where each element of the array is a string (possibly the empty string). We refer to a column as a \textit{field}, and denote the $k^{th}$ field $c^k$. A row is referred to as a \textit{record}, with $r_i$ denoting the $i^{th}$ record of the data set. An element of the array is referred to as an \textit{entry}, denoted $e_{i,k}$ (referring to the $i^{th}$ entry in the $k^{th}$ field). Each entry can contain multiple features where a \textit{feature} is a string of characters. There is significant freedom in choosing how to divide the string which makes up entry $e_{i,k}$ into multiple features. In our implementations in this paper we compare two different methods: (1) cutting the string at white spaces and (2) dividing the string into \textit{$N$-grams}. For example, consider an entry $e_{i,k}$ which is made up of the string ``Albert Einstein''. Following method (1) this entry has two features: ``Albert'' and `'Einstein''. Method (2), the $N$-gram representation, creates features $f^k_1,\ldots, f^k_L$, corresponding to all possible substrings of $e_{i,k}$ containing $N$ consecutive characters (if an entry contains $N$ characters or fewer, the full entry is considered to be a single token). Hence $L$ is equal to the length of the string minus $(N-1)$. In our example, if we use $N=3$, $e_{i,k}$ contains 13 features. Ordered alphabetically (with white space `` '' preceding ``A''), the features are
\begin{align*} f^k_1 &=  \text{`` Ei''}, \
f^k_2 =  \text{``Alb''}, \
f^k_3 =  \text{``Ein''}, \
f^k_4 =  \text{``ber''}, \\
f^k_5 &=  \text{``ein''}, \
f^k_6 =  \text{``ert''}, \
f^k_7 =  \text{``ins''}, \
f^k_8 =  \text{``lbe''}, \\
f^k_9 &=  \text{``nst''}, \
f^k_{10} =  \text{``rt ''}, \
f^k_{11} =  \text{``ste''}, \
f^k_{12} =  \text{``t E''},\\
f^k_{13} &=  \text{``tei''}.
\end{align*}
In our applications we remove any $N$-grams that consist purely of white spaces.

When discussing our results we will specify where we have used method (1) and where we have used method (2), by indicating if we have used {\em word features} or {\em $N$-gram features} respectively. 

For each field we create a dictionary of all features in that field and then remove stop words or words that are irrelevant, such as ``and'', ``the'', ``or'', ``None'', ``NA'', or `` '' (the empty string). We refer to such words collectively as ``stop words'' (as is common in practice) and to this reduced dictionary as the \textit{set of features}, $f^k$, where: 
\begin{equation*}
f^k := \big( f_{1}^k, f_{2}^k, \ldots, f_{m-1}^k, f_{m}^k\big), 
\end{equation*}
if the $k^{\text{th}}$ field contains $m$ features. This reduced dictionary represents an ordered set of unique features found in field $c^k$.

Note that $m$, the number of features in $f^k$, depends on $k$, since a separate set of features is constructed for each field. To keep the notation as simple as possible, we will not make this dependence explicit in our notation. Since, in this paper, $m$ is always used in the context of a given, fixed $k$, this should not lead to confusion.

We will write $f_j^k \in e_{i,k}$ if the entry $e_{i,k}$ contains the feature $f_j^k$. Multiple copies of the same feature can be contained in any given entry. This will be explored further in Section~\ref{sec:TFIDF}. Note that an entry can be ``empty" if it only contains stop words, since those are not included in the set of features $f^k$.

We refer to a subset of records as a \textit{cluster} and denote it $R = \{r_{t_1}, \ldots, r_{t_p}\}$ where each $t_i \in \{1, 2, \ldots n\}$ is the index of a record in the data set.

The duplicate detection problem can then be stated as follows: Given a data set containing duplicate records, find clusters of records that represent a single entity, i.e., subsets containing those records that are duplicates of each other. \textit{Duplicate records}, in this sense, are not necessarily identical records but can also be `near identical' records. They are allowed to vary due to spelling errors or missing entries.

\section{Related methods}\label{sec:existingmethods}

Numerous algorithms for duplicate detection exist, including various probabilistic methods \cite{jaro95}, string comparison metrics \cite{jaro_original,winkler_original}, feature frequency methods \cite{salton1988term}, and hybrid methods \cite{soft_tfidf_original_cohen}. There are many other proposed methods for data matching, record linkage and various stages of data cleaning, that have a range of success in specific applications but also come with their own limitations and drawbacks.  Surveys of various duplicate detection methods can be found in \cite{elmagarmid2007duplicate,Allison2009,Hassanzadeh2009,AbrahamKanmani2014,Ramya2017}.

Probabilistic rule based methods, such as Fellegi-Sunter based models \cite{winkler_original}, are methods that attempt to learn features and rules for record matching using conditional probabilities, however, these are highly sensitive to the assumed model which is used to describe how record duplicates are distributed across the database and become completely infeasible at large scale when comparing all pairs. Other rule based approaches such as \cite{Tamilselvi2011} attempt to create a set of rules that is flexible enough to deal with different types of data sets.

Privacy-preserving record matching techniques \cite{hall2010privacy,scannapieco2007privacy}, based on hash encoding, are fast and scalable, but can only handle exact matching (single character differences or small errors in input result in completely different hash codes); approximate matching based methods are often possible but typically not scalable.  

Collective record matching techniques \cite{nuray2013adaptive,fu2012multiple} have been proposed that match records across multiple databases, using a graph based on similarity of groups of entities.  These methods have shown promise in some applications where entity relationships are identifiable (such as sharing the same address or organization), but direct applications are limited and are currently not generalizable or scalable.   

Unsupervised or supervised techniques \cite{friedman2001elements} can also be used directly, using records as features, but in most applications labeled data does not exist for training or evaluation.  Additionally, standard testing data sets, used for comparing methods, are extremely limited and weakly applicable to most applications. Some techniques are developed specifically to deal with hierarchical data, such as XML data \cite{Leitaoetal2013,AbrahamKanmani2014}. We do not consider that situation here.

For larger data sets a prefix filtering \cite{Xiao2011}, blocking \cite{DeVries2011,Draisbachetal2011,Papadakis2011,Papadakis2013} or windowing \cite{Draisbachetal2011,BanoAzam2015,RameshKannanetal2016} step can be used. Such methods can be seen as a preprocessing step which identifies records which are not likely to be duplicates, such that the pairwise feature similarity does only need to be computed for those features that co-appear in likely duplicates. A survey of various such indexing methods is given in \cite{Christen2012}. We did not include an indexing step in our experiments in this paper, so that our experiments are run without excluding any record pairings a priori, but they can be incorporated into our method

Pay-as-you-go \cite{Whang2013} or progressive duplicate detection methods \cite{Papenbrock2015,RameshKannanetal2016} have been developed for applications in which the duplicate detection has to happen in limited time on data which is acquired in small batches or in (almost) real-time \cite{Layek2016}. In our paper we consider the situation in which we have all data available from the start.

In \cite{Bilenko2003} the authors suggest to use trainable similarity measures that can adapt to different domains from which the data originate. In this paper we develop our method using given similarity measures, such that our method is applicable in the absence of training data.

In the remainder of this section we present in more detail those methods which are related to the proposed method we introduce in Section~\ref{sec:newmethod}. We review both the \textit{Jaro} and \textit{Jaro-Winkler} string metrics, the feature frequency based \textit{term frequency-inverse document frequency} (TF-IDF) method, and the hybrid \textit{soft TF-IDF} method.

\subsection{Character-based similarity: Jaro and Jaro-Winkler}\label{sec:Jaro}
Typographical variations are a common cause of duplication among string data, and the prevalence of this type of error motivates string comparison as a method for duplicate detection. The Jaro distance \cite{jaro_original} was originally devised for duplicate detection in government census data and modified by Winkler \cite{winkler_original} to give more favorable similarities to strings with matching prefixes. This latter variant is now known as the Jaro-Winkler string metric and has been found to be comparable empirically with much more complex measures \cite{soft_tfidf_original_cohen}. Despite their names, neither the Jaro distance, nor the Jaro-Winkler metric, are in fact distances or metrics in the mathematical sense, since they do not satisfy the triangle inequality, and exact matches have a score of 1, not 0. Rather, they can be called similarity scores.

\begin{figure}[htp]
\centering
\includegraphics[width=0.3\textwidth]{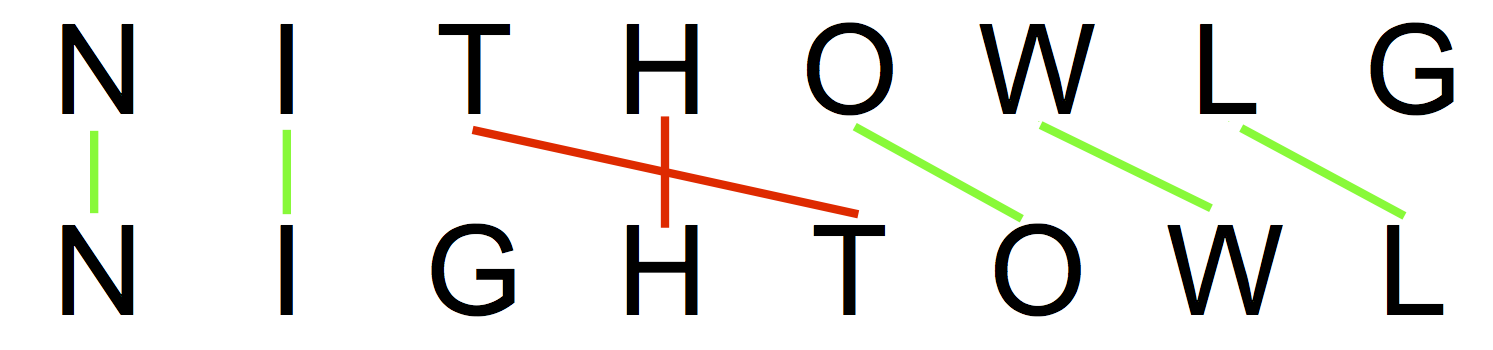}
\caption{Example of a comparison of two features in the computation of the Jaro distance, with character window size $W=4$. The example has 7 matching character pairs, 2 of which are transposition pairs, represented by the red lines. The green lines indicate matching pairs that are not transpositions. Notice that ``G'' is not considered a matching character as ``G'' in ``NITHOWLG'' is the 8th character while ``G'' in ``NIGHTOWL'' is the 3rd character, which is out of the $W=4$ window for this example. Here, $J = \frac{1}{3}(\frac{7}{8}+\frac{7}{8} + \frac{7-1}{7})=0.869$.}
\label{fig:jaro}
\end{figure}

To define the Jaro-Winkler metric, we must first define the Jaro distance. For two features $f_i^k$ and $f_j^k$, we define the \textit{character window size}
\begin{equation*}
W_{i,j}^k := \bigg\lfloor \dfrac{\min(|f_i^k|,|f_j^k|)}{2} \bigg\rfloor,
\end{equation*}
where $|f_i^k|$ is the length of the string $f_i^k$, i.e., the number of characters in $f_i^k$ counted according to multiplicity.
The $l^{th}$ character of the string $f_i^k$ is said to \textit{match} the ${l^\prime}^{th}$ character of $f_j^k$, if both characters are identical and $l-W_{i,j}^k \leq l' \leq l+W_{i,j}^k$. Let $M$ be the number of characters in string $f_i^k$ that match with characters in string $f_j^k$ (or, equivalently, the number of characters in $f_j^k$ that match with characters in $f_i^k$), let $(a_1, \ldots, a_M)$ be the matched characters from $f_i^k$ in the order they appear in the string $f_i^k$, and let $(b_1, \ldots, b_M)$ be the matched characters from $f_j^k$ in order. Then $t$ is defined to be half the number of \textit{transpositions} between $f_i^k$ and $f_j^k$, i.e., half the number of indices $l\in \{1, \ldots, M\}$ such that $a_l \neq b_l$. Each such pair $(a_l, b_l)$ is called a \textit{transposition pair}. Now the \textit{Jaro distance} \cite{jaro_original} $J(f_i^k, f_j^k)$ is defined as
\begin{align*}
&J(f_i^k, f_j^k) :=\\
&\hspace{0.1cm} \begin{cases} \dfrac{1}{3}\bigg (\dfrac{M}{|f_i^k|}+\dfrac{M}{|f_j^k|}+\dfrac{M-t}{M} \bigg ),   &\text{if } M\neq 0, \\
0, & \text{if } M=0. \end{cases}
\end{align*}
Fig.~\ref{fig:jaro} shows an example of transpositions and matching character pairs.

The Jaro-Winkler metric, $\mat{JW}(f_i^k,f_j^k)$, modifies the original Jaro distance by giving extra weight to matching prefixes. It uses a fixed \textit{prefix factor} $p$ to give a higher similarity score to features that start with the same characters. Given two features $f_i^k$ and $f_j^k$, the \textit{Jaro-Winkler metric} is
\begin{equation}\label{eq:JaroWinkler}
\mat{JW}(f_i^k, f_j^k) := J(f_i^k,f_j^k)+p\, \ell_{i,j}\left(1-J(f_i^k,f_j^k)\right),
\end{equation}
where $J(f_i^k, f_j^k)$ is the Jaro distance between two features $f_i^k$ and $f_j^k$, $p$ is a given prefix factor, and $\ell_{i,j}$ is the number of prefix characters in $f_i^k$ that are the same as the corresponding prefix characters in $f_j^k$ (i.e., the first $\ell_{i,j}$ characters in $f_i^k$ are the same as the first $\ell_{i,j}$ characters in $f^j_k$ and the $(\ell_{i,j}+1)^{\text{th}}$ characters in both features differ). When we want to stress that, for fixed $k$, $\mat{JW}(f_i^k, f_j^k)$ is an element of a matrix, we write $\mat{JW}_{i,j}^k := \mat{JW}(f_i^k, f_j^k)$, such that $\mat{JW}^k \in \R^{m\times m}$.

In Winkler's original work he set $p=0.1$ and restricted $\ell_{i,j} \leq 4$ (even when prefixes of five or more characters were shared between features) \cite{winkler_original}. We follow the same parameter choice and restriction in our applications in this paper. So long as $p\, \ell_{i,j} \leq 1$ for all $i, j$, the Jaro-Winkler metric ranges from 0 to 1, where 1 indicates exact similarity between two features and 0 indicates no similarity between two features.

In Fig.~\ref{fig:jaro} we have $\ell=2$, as both features have identical first and second characters, but not a matching third character. This leads to $\mat{JW} = 0.869+0.1 \cdot 2 \cdot (1-0.869) = 0.895$.

Because we remove stop words and irrelevant words from our set of features, it is possible for an entry $e_{i,k}$ to contain a feature that does not appear in $f^k$. If a feature $\tilde f \in e_{i,k}$ does not appear in the dictionary $f^k$, we set, for all $f_q^k \in f^k$, $\mat{JW}(f_q^k, \tilde f) := 0$. We call such features $\tilde f$ \textit{null features}.

\vspace{.5cm}
\begin{algorithm}
\KwData{$c^k$, an $n \times 1$ array of text}
\KwResult{$\mat{JW}^k\in \mathbb{R}^{m \times m}$}
Create the set of features $f^k = (f_1^k, \ldots, f_m^k)$\\
\For{each pair of features $(f_i^k, f_j^k)$}{
	\emph{Compute Jaro distance $J_{i,j} = J(f_i^k, f_j^k)$} \\
\emph{Compute Jaro-Winkler similarity $\mat{JW}_{i,j}^k =  \begin{cases}J_{i,j}+p\, \ell_{i,j}(1-J_{i,j}), &\text{if neither feature } \\&f_i^k \text{ or } f_j^k \text{ is a}
\\ &\text{null feature}, \\ 0, &\text{else} \end{cases}$}
}
\caption{Jaro-Winkler Algorithm}
\end{algorithm}

\subsection{Feature-based similarity: TF-IDF}\label{sec:TFIDF}

Another approach to duplicate detection, generally used in big data record matching, looks at similar distributions of features across records. This feature based method considers entries to be similar if they share many of the same features, regardless of order; this compensates for errors such as changes in article usage and varying word order (e.g. ``The Bistro'', ``Bistro, The'', or ``Bistro''), as well as the addition of information (e.g. ``The Bistro'' and ``The Bistro Restaurant''). 

This form of duplicate detection is closely related to vector space models of text corpora \cite{vsm_original}, where a body of text is represented as a vector in some word vector space. The dimension of the space is the number of relevant words (other words are assumed to be meaningless), and, for a given record, each element of the vector representation is the frequency with which a word appears in the entry. (It should be noted that these models also disregard word order.) A more powerful extension of these models is the term frequency-inverse document frequency (TF-IDF) scheme \cite{salton1988term}. This scheme reweighs different features based on their frequency in a single field as well as in an entry.

Using the reduced set of features, $f^k$, we create the term frequency and inverse document frequency matrices. We define the \textit{term frequency matrix} for the $k^{th}$ field, $\mat{TF}^k \in \mathbb{R}^{n \times m}$, such that $\mat{TF}_{i,j}^k$ is the number of times the feature $f_j^k$ appears in the entry $e_{i,k}$ (possibly zero). A row of $\mat{TF}^k$ represents the frequency of every feature in an entry.

Next we define the diagonal \textit{inverse document frequency matrix} $\mat{IDF}^k \in \mathbb{R}^{m \times m}$ with diagonal elements\footnote{We use $\log$ to denote the natural logarithm in this paper.}
\begin{equation*}
\mat{IDF}_{i,i}^k := \log\dfrac{n}{|\{e \in c^k : f_i^k \in e\}|},
\end{equation*}
where $|\{e \in c^k: f_i^k \in e\}|$ is the number of entries\footnote{By the construction of our set of features in Section~\ref{sec:terminology}, this number of entries is always positive.} in field $c^k$ containing feature $f_i^k$, and where $n$ is the number of records
in the data set. The matrix $\mat{IDF}^{k}$ uses this number of entries in the field which contain a given feature to give this feature a more informative weight. 
The issue when using term frequency only, is that it gives features that appear frequently a higher weight than rare features. The latter often are empirically more informative than common features, since a feature that occurs frequently in many entries is unlikely to be a good discriminator.

The resulting weight matrix for field $k$ is then defined with a logarithmic scaling for the term frequency as\footnote{Note that, following \cite{soft_tfidf_original_cohen}, we use a slightly different logarithmic scaling, than the more commonly used $\mat{TFIDF}_{i,j}^k = \big(\log(\mat{TF}_{i,j}^k) + 1\big)\mat{IDF}_{i,i}^k$, if $\mat{TF}_{i,j}^k\neq 0$, and $\mat{TFIDF}_{i,j}^k = 0$, if $\mat{TF}_{i,j}^k=0$. This avoids having to deal with the case $\mat{TF}_{i,j}^k=0$ separately. The difference between $\log(\mat{TF}_{i,j}^k) + 1$ and $\log(\mat{TF}_{i,j}^k + 1)$ is bounded by $1$ for $\mat{TF}_{i,j}^k\geq 1$.}
\begin{equation}\label{eq:TFIDF}
\mat{TFIDF}^k := N^k \log(\mat{TF}^k + \textbf{1})\mat{IDF}^k,
\end{equation}
where $\textbf{1}$ is an $n \times m$ matrix of ones, the $\log$ operation acts on each element of $\mat{TF}^k + \textbf{1}$ individually, and $N^k\in \R^{n\times n}$ is a diagonal normalization matrix such that each nonzero row of $\mat{TFIDF}^k$ has unit $\ell^1$ norm\footnote{Here we deviate from \cite{soft_tfidf_original_cohen}, in which the authors normalize by the $\ell^2$ norm. We do this so that later in equation \eqref{eq:sTFIDF}, we can guarantee that the soft TF-IDF values are upper bounded by 1.}. The resulting matrix has dimension $n \times m$. Each element $\mat{TFIDF}^k_{i,j}$ represents the weight assigned to feature $j$ in field $k$ for record $i$. Note that each element is nonnegative.

\vspace{.5cm}
\begin{algorithm}
\KwData{$c^k$, an $n \times 1$ array of text}
\KwResult{$\mat{TFIDF}^k \in \mathbb{R}^{n \times m}$}
\emph{Create the set of features $f^k = (f_1^k, \ldots, f_m^k)$}\\
\For{each pair of features $(f_i^k, f_j^k)$}{
	\emph{Compute term frequency $\mat{TF}_{i,j}^k$}
}
\For{each feature $f_i^k$}{
	\emph{Compute inverse document frequency $\mat{IDF}_{i,i}^k$}
}
\emph{Initialize $\mat{TFIDF}^k = \log (\mat{TF}^k+\textbf{1})\mat{IDF}^k$}\\
\emph{Normalize rows of $\mat{TFIDF}^k$ to have unit $\ell^1$ norm}
\caption{TF-IDF Algorithm}
\end{algorithm}

\subsection{Hybrid similarity: soft TF-IDF}\label{sec:sTFIDF}

The previous two methods concentrate on two different causes of record duplication, namely typographical error and varying word order. It is easy to imagine, however, a case in which both types of error occur; this leads us to a third class of methods which combine the previous two. These \textit{hybrid methods} measure the similarity between entries using character similarity between their features as well as weights of their features based on importance. Examples of these hybrid measures include the extended Jacard similarity and the Monge-Elkan measure \cite{introdupdet}. In this section we will discuss another such method, soft TF-IDF \cite{soft_tfidf_original_cohen}, which combines TF-IDF with a character similarity measure. In our method, we use the Jaro-Winkler metric, discussed above in Section~\ref{sec:Jaro}, as the character similarity measure in soft TF-IDF.

For $\theta \in [0,1)$, let $S^k_{i,j}(\theta)$ be the set of all index pairs $(p,q) \in \R^{m\times m}$ such that $f_p^k\in e_{i,k}$, $f_q^k \in e_{j,k}$, and $\mat{JW}(f_p^k,f_q^k) > \theta$, where $\mat{JW}$ is the Jaro-Winkler similarity metric from \eqref{eq:JaroWinkler}. The \textit{soft TF-IDF similarity score} between two entries $e_{i,k}$ and $e_{j,k}$ in field $c^k$ is defined as
\begin{align}
&\mat{sTFIDF}_{i,j}^k :=\label{eq:sTFIDF}\\
& \begin{cases}
\hspace{-0.1cm}\underset{(p,q)\in S^k_{i,j}(\theta)}\sum \hspace{-0.3cm} \mat{TFIDF}_{i,p}^k \cdot \mat{TFIDF}_{j,q}^k \cdot \mat{JW}^k_{p,q}, &\text{if } i\neq j,\\
1, &\text{if } i=j.
\end{cases}\notag
\end{align}
The parameter $\theta$ allows for control over the similarity of features, removing entirely pairs that do not have Jaro-Winkler similarity above a certain threshold. The results presented in this paper are all obtained with $\theta = 0.90$.

The soft TF-IDF similarity score between two entries is high if they share many similar features, where the similarity between features is measured by the Jaro-Winkler metric and the contribution of each feature is weighted by its TF-IDF score. If we contrast the soft TF-IDF score with the TF-IDF score described in Section~\ref{sec:TFIDFinstead} below, we see that the latter only uses those features which are exactly shared by both entries, whereas the former also incorporates contributions from features that are very similar (but not exactly the same). This means that the soft TF-IDF score allows for high similarity between entries in the presence of both misspellings and varying word (or feature) order more so than the TF-IDF score does.

Note from \eqref{eq:sTFIDF} that for all $i$, $j$, and $k$, we have $\mat{sTFIDF}_{i,j}^k\in [0,1]$. The expression for the case $i\neq j$ does not necessarily evaluate to $1$ in the case $i=j$. Therefore we explicitly included $\mat{sTFIDF}_{i,i}^k=1$ as part of the definition, since this is a reasonable property for a similarity measure to have. Luckily, these diagonal elements of $\mat{sTFIDF}^k$ will not be relevant in our method, so the $i=j$ part of the definition is more for definiteness and computational ease\footnote{The values of the diagonal elements are not relevant theoretically, because any record is always a `duplicate' of itself and trivially will be classified as such, i.e. each record will be clustered in the same cluster as itself. However, if the diagonal elements are not set to have value $1$, care must be taken that this does not influence the numerical implementation.}, than out of strict necessity for our method.

In practice, this method's computational cost is greatly reduced by vectorization.  Let $M^{k,\theta} \in \mathbb{R}^{m \times m}$ be the Jaro-Winkler similarity matrix defined by
\begin{equation*}
M^{k,\theta}_{p,q} := \begin{cases}  	\mat{JW}(f_p^k, f_q^k),	&\text{if } \mat{JW}(f_p^k, f_q^k) \geq \theta, \\
0, & \text{if } \mat{JW}(f_p^k, f_q^k) < \theta. \end{cases}
\end{equation*}

The soft TF-IDF similarity for each $(i,j)$ pairing ($i\neq j$) can then be computed as
\[ \mat{sTFIDF}_{i,j}^k = \sum_{\mathclap{p,q = 1}}^m \left[ \left({\mat{TFIDF}_i^k}^T \mat{TFIDF}_j^k\right) * M^{k,\theta}\right]_{p,q}, \]
where TFIDF$_i^k$ denotes the $i^{th}$ row of the TF-IDF matrix of field $c^k$ and $*$ denotes the Hadamard product (i.e. the element-wise product). We can further simplify this using tensor products. Let $\overline{M}^{k,\theta}$ denote the vertical concatenation of the rows of $M^{k,\theta}$.
\begin{equation*}
\overline{M}^{k,\theta}= \begin{bmatrix} 	{M_1^{k,\theta}}^T\\
								{M_2^{k,\theta}}^T \\
								\vdots \\
								{M_m^{k,\theta}}^T \end{bmatrix}
\end{equation*}
where $M_i^{k,\theta}$ is the $i^{th}$ row of $M^{k,\theta}$. We then have
\begin{equation*}
\mat{sTFIDF}_{i,j}^k = (\mat{TFIDF}_i^k \otimes \mat{TFIDF}_j^k) * \overline{M}^{k,\theta},
\end{equation*}
if $i\neq j$. Here $\otimes$ is the Kronecker product. Finally we set the diagonal elements $\mat{sTFIDF}_{i,i}^k = 1$.

\vspace{.5cm}
\begin{algorithm}
\KwData{$\mat{JW}^k \in \mathbb{R}^{m \times m}$, $\mat{TFIDF}^k \in \mathbb{R}^{n\times m}$, $\theta$}
\KwResult{$\mat{sTFIDF}^k \in \mathbb{R}^{n \times n}$}
\emph{Create the set of features $f^k = (f_1^k, \ldots, f_m^k)$}\\
\For{each pair of features $(f_i^k, f_j^k)$}{
	\emph{Compute the thresholded Jaro-Winkler matrix $M^{k, \theta}_{i,j}$}
}
\emph{Vertically concatenate rows of $M^{k, \theta}$: \,$\overline{M}^{k, \theta} = [{M_1^{k, \theta}}^T; {M_2^{k, \theta}}^T; \ldots ;{M_m^{k, \theta}}^T ]$} \\
\For{each pair of entries $(e_{i,k}, e_{j,k})$ in field $c^k$ }{
\emph{Compute soft TF-IDF for $i\neq j$: $\mat{sTFIDF}^{\,k}_{i,j} = (\mat{TFIDF}_i^{\,k} \otimes \mat{TFIDF}_j^{\,k}) * \overline{M}^{k, \theta}$}
}
\emph{Set the diagonal elements $\mat{sTFIDF}_{i,i}^k = 1$}
\caption{soft TF-IDF Algorithm}
\end{algorithm}

The TF-IDF and Jaro-Winkler similarity matrices are typically sparse. This sparsity can be leveraged to reduce the computational cost of the soft TF-IDF method as well.

The soft TF-IDF scores above are defined between entries for a single field. For each pair of records we produce a \textit{composite similarity score} $\mat{ST}_{i,j}$ by adding their soft TF-IDF scores over all fields:

\begin{equation}\label{eq:ST}
\mat{ST}_{i,j} :=  \sum_{k=1}^{a} \mat{sTFIDF}^k_{i,j}.
\end{equation}
Hence $\mat{ST} \in \mathbb{R}^{n \times n}$ and $\mat{ST}_{i,j}$ is the score between the $i^{th}$ and $j^{th}$ records. Remember that $a$ is the number of fields in the data set, thus each composite similarity score $\mat{ST}_{i,j}$ is a number in $[0,a]$.

For some applications it may be desirable to let some fields have a greater influence on the composite similarity score than others. In the above formulation this can easily be achieved by replacing the sum in \eqref{eq:ST} by a weighted sum:
\[
\mat{ST}^w_{i,j} :=  \sum_{k=1}^{a} w_k\, \mat{sTFIDF}^k_{i,j},
\]
for positive weights $w_k \in \R$, $k\in \{1, \ldots, a\}$. If the weights are chosen such that $\sum_{k=1}^a w_k \leq a$, then the \textit{weighted composite similarity scores} $\mat{ST}^w_{i,j}$ take values in $[0,a]$, like $\mat{ST}_{i,j}$. In this paper we use the unweighted composite similarity score matrix $\mat{ST}$.

\subsection{Using TF-IDF instead of soft TF-IDF}\label{sec:TFIDFinstead}

In our experiments in Section~\ref{sec:Applications} we will also show results in which we use TF-IDF, not soft TF-IDF, to compute similarity scores. This can be achieved in a completely analogous way to the one described in Section~\ref{sec:sTFIDF}, if we replace $\mat{JW}^k_{p,q}$ in \eqref{eq:sTFIDF} by the Kronecker delta 
$
\delta_{p,q} := \begin{cases} 1, & \text{if } p=q,\\ 0, & \text{otherwise}.\end{cases}
$
The dependency on $\theta$ disappears and we get
\begin{equation}\label{eq:sTFIDFforTFIDF} 
\mat{sTFIDF}_{i,j}^k :=\begin{cases}
\overset{m}{\underset{p=1}\sum}\left(\mat{TFIDF}_{i,p}^k\right)^2, &\text{if } i\neq j,\\
1, &\text{if } i=j.
\end{cases}
\end{equation}
Note that the values for $i\neq j$ correspond to the off-diagonal values in the matrix $\mat{TFIDF}^k \big(\mat{TFIDF}^k\big)^T \in \mathbb{R}^{n\times n}$, where $\mat{TFIDF}^k$ is the TF-IDF matrix from \eqref{eq:TFIDF} and the superscript $T$ denotes the matrix transpose\footnote{Our choice to normalize the rows of $\mat{TFIDF}^k$ by their $\ell^1$ norms instead of their $\ell^2$ norms means that the diagonal elements of $\mat{TFIDF}^k \big(\mat{TFIDF}^k\big)^T$ are not necessarily equal to $1$.}.

We used the same notation for the matrices in \eqref{eq:sTFIDF} and \eqref{eq:sTFIDFforTFIDF}, because all the other computations, in particular the computation of the composite similarity score in \eqref{eq:ST} which is used in the applications in Section~\ref{sec:Applications}, follow the same recipe when using either matrix. Where this is of importance in this paper, it will be clear from the context if $\mat{ST}$ has been constructed using the soft TF-IDF or TF-IDF similarity scores.

\section{The proposed methods}\label{sec:newmethod}

We extend the soft TF-IDF method to address two common situations in duplicate detection: sparsity due to missing entries and large numbers of duplicates. For data sets with only one field, handling a missing field is a non-issue; a missing field is irreconcilable, as no other information is gathered. In a multi-field setting, however, we are faced with the problem of comparing partially complete records. Another issue is that a record may have more than one duplicate. If all entries are pairwise similar we can easily justify linking them all, but in cases where one record is similar to two different records which are dissimilar to each other the solution is not so clear cut.

\begin{figure}
\centering
\includegraphics[width=0.5\textwidth]{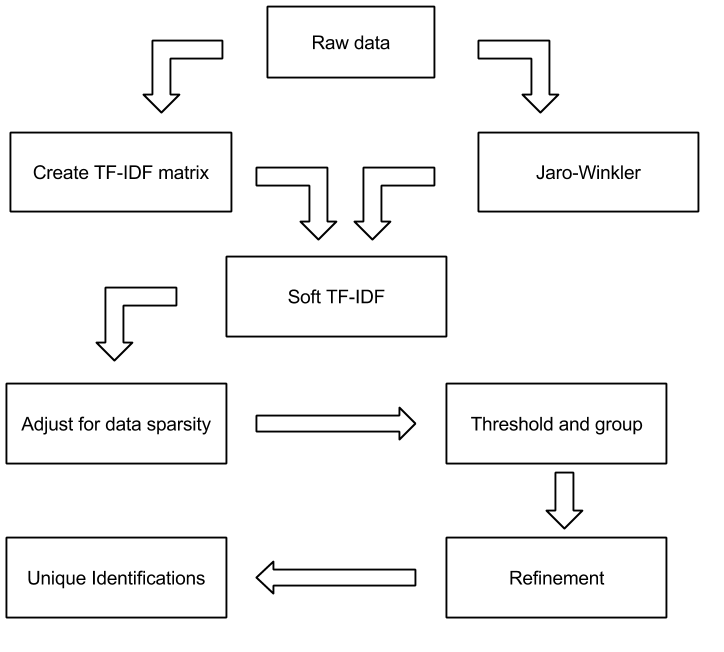}
\caption{An outline of our method for duplicate detection}
\label{fig:method}
\end{figure}

Fig.~\ref{fig:method} shows an outline of our method. First we use TF-IDF to assign weights to features that indicate the importance of that feature in an entry. Next, we use soft TF-IDF with the Jaro-Winkler metric to address spelling inconsistencies in our data sets. After this, we adjust for sparsity by taking into consideration whether or not a record has missing entries. Using the similarity matrix produced from the previous steps, we threshold and group records into clusters. Lastly, we refine these groups by evaluating how clusters break up under different conditions.

\subsection{Adjusting for sparsity}\label{sec:adjusting}
A \textit{missing entry} is an entry that is either entirely empty from the start or one that contains only null features and thus ends up being empty for our purposes. Here, we assume that missing entries do not provide any information about the record and therefore cannot aid us in determining whether two records should be clustered together (i.e. labeled as probable duplicates). In \cite{tromp2006record}, \cite{winkler_original}, and \cite{allison2005imputation}, records with missing entries are discarded, filled in by human fieldwork, and filled in by an expectation-maximization (EM) imputation algorithm, respectively. For cases in which a large number of entries are missing, or in data sets with a large number of fields such that records have a high probability of missing at least one entry, these first two methods are impractical. Furthermore, the estimation of missing fields is equivalent to unordered categorical estimation. In fields where a large number of features are present (i.e. the set of features is large), estimation by an EM scheme becomes computationally intractable \cite{pigott2001review} \cite{Winkler02methodsfor} \cite{horton2007much}. Thus, a better method is required.

Leaving the records with missing entries in our data set, both TF-IDF and Jaro-Winkler remain well defined, allowing (soft) TF-IDF schemes to proceed. However, because the Jaro-Winkler metric between a null feature and any other feature is $0$, the soft TF-IDF score between a missing entry and any other entry is $0$. This punishes sparse records in the composite soft TF-IDF similarity score matrix $\mat{ST}$. Even if two records have the exact same entries in fields where both records do not have missing entries, their missing entries deflate their composite soft TF-IDF similarity. Consider the following example using two records (from a larger data set containing $n>2$ records) and three fields: [``Joe Bruin'', `` '', ``male''] and [``Joe Bruin', ``CA'', `` '']. The two records are likely to represent a unique entity ``Joe Bruin'', but the composite soft TF-IDF score between the two records is on the lower end of the similarity score range (1 out of a maximum of 3) due to the missing entry in the second field for the first record and the missing entry in the third field for the second record. The issue described above for the soft TF-IDF method is also present for the TF-IDF method described in Section~\ref{sec:TFIDFinstead}.

To correct for this, we take into consideration the number of mutually present (not missing) entries in the same field for two records.
This can be done in a vectorized manner to accelerate computation. Let B be the $n \times a$ binary matrix defined by
\begin{equation*}
B_{i,k} := \begin{cases}  0,    &\text{if } e_{i,k} \text{ is a missing entry}, \\
1,  &\text{otherwise}. \end{cases}
\end{equation*}

This is a binary mask of the data set, where 1 denotes a non-missing entry (with or without error), and 0 denotes a missing entry. In the product $BB^T \in \mathbb{R}^{n \times n}$, each $(BB^T)_{i,j}$ is the number of ``shared fields'' between records $r_i$ and $r_j$, i.e. the number of fields $c^k$ such that both $e_{i,k}$ and $e_{j,k}$ are non-missing entries. Our \textit{adjusted (soft) TF-IDF similarity score} is given by
\begin{equation}
\mat{adjST}_{i,j} :=  \begin{cases}
\frac{\mat{ST}_{i,j}}{(BB^T)_{i,j}}, &\hspace{-0.1cm}\text{if } i\neq j \text{ and }  (BB^T)_{i,j} \neq 0,\\
0, &\hspace{-0.1cm}\text{if } i\neq j \text{ and }  (BB^T)_{i,j} = 0,\\
1, &\hspace{-0.1cm}\text{if } i=j.
\end{cases}
\label{eq:adjST}
\end{equation}
Remembering that $\mat{JW}(f_p^k,f_q^k)=0$ if $f_p^k$ is a null feature or $f_q^k$ is a null feature, we see that, if $e_{i,k}$ is a missing entry or $e_{j,k}$ is a missing entry, then the set $S^k_{i,j}(\theta)$ used in \eqref{eq:sTFIDF} is empty (independent of the choice of $\theta$) and thus $\mat{sTFIDF}_{i,j}^k=0$. The same conclusion is true in \eqref{eq:sTFIDFforTFIDF} since the $i^{\text{th}}$ or $j^{\text{th}}$ row of $\mat{TFIDF}^k$ consists of zeros in that case. Hence, we have that, for all $i, j$ ($i\neq j$), $(\mat{ST})_{i,j} \in [0,(BB^T)_{i,j}]$ (which refines our earlier result that $(\mat{ST})_{i,j} \in [0,a]$) and thus $(\mat{adjST})_{i,j} \in [0,1]$.

In the event that there are records $r_i$ and $r_j$ such that $(BB^T)_{i,j}=0$, it follows that $\mat{ST}_{i,j}=0$. Hence it makes sense to define $\mat{adjST}_{i,j}$ to be zero in this case. In the data sets we will discuss in Section~\ref{sec:Applications}, no pair of records was without shared fields. Hence we can use the shorthand expression $\mat{adjST} = \mat{ST} \oslash BB^T$ for our purposes in this paper\footnote{Since we defined the inconsequential diagonal entries to be $\mat{sTFIDF}_{i,i}^k = 1$ in \eqref{eq:sTFIDF} and \eqref{eq:sTFIDFforTFIDF}, it could be that $(\mat{ST})_{i,i} > (BB^T)_{i,i}$ for some $i$, which is why we explicitly defined $(\mat{adjST})_{i,i} = 1$ in \eqref{eq:adjST} for consistency with the other values. Since the diagonal values will play no role in the eventual clustering this potential discrepancy between \eqref{eq:adjST} and $\mat{adjST} = \mat{ST} \oslash BB^T$ is irrelevant for our purposes.}, where $\oslash$ denotes element-wise division.

\vspace{.5cm}
\begin{algorithm}
\KwData{$\mat{sTFIDF}^k \in \mathbb{R}^{n \times n}$ for $k \in \{1, \ldots, a\}$, $D$ an $n \times a$ array of text}
\KwResult{$\mat{adjST} \in \mathbb{R}^{n \times n}$}
\For{each entry $e_{i,k}$ in each field $c^k$ of $D$}{
	\emph{Compute $B_{i,k}$}\\
}
\emph{Initialize} $\mat{ST} = \sum_k \mat{sTFIDF}^k$\\
\emph{Adjust $\mat{ST}$ for sparsity:} $\mat{adjST} = \mat{ST} \oslash BB^T$\\
\caption{Adjusting for Sparsity}
\end{algorithm}

Instead of the method proposed above to deal with missing data, we can also perform data imputation to replace the missing data with a ``likely candidate'' \cite{KimGolubPark2005,KimGolubPark2006,Allison2009,Huisman2009,larose2014discovering,Watadaetal2016}. To be precise, before computing the matrix $B$, we replace each missing entry $e_{i,k}$ by the entry which appears most often in the $k^{\text{th}}$ field\footnote{We use the mode, rather than the mean, because all our data is either textual or, when numeric, it is ordinal, rather than cardinal, such as in the case of social security numbers.}. In case of a tie, we choose an entry at random among all the entries with the most appearances (we choose this entry once per field, such that each missing entry in a given field is replaced by the same entry). For a clean comparison, we still compute the matrix $B$ (which has now no $0$ entries) and use it for the normalization in \eqref{eq:adjST}. The rest of our method is then implemented as usual. We report the results of this comparison in Section~\ref{sec:imputation}.

\subsection{Thresholding and grouping}\label{sec:TGS}

The similarity score $\mat{adjST}_{i,j}$ gives us an indication of how similar the records $r_i$ and $r_j$ are. If $\mat{adjST}_{i,j}$ is close to $1$, then the records are more likely to represent the same entity. Now, we present our method of determining whether a set of records are duplicates of each other based on $\mat{adjST}$. There exist many clustering methods that could be used to accomplish this goal. For example, \cite{monge1997efficient} considers this question in the context of duplicate detection. For simplicity, in this paper we restrict ourselves to a relatively straightforward thresholding procedure, but other methods could be substituted in future implementations. We call this the \textit{thresholding and grouping step} (TGS).

The method we will present below is also applicable to clustering based on other similarity scores. Therefore it is useful to present it in a more general format. Let $\mat{SIM} \in \R^{n\times n}$ be a matrix of similarity scores, i.e., for all $i, j$, the entry $\mat{SIM}_{i,j}$ is a similarity score between the records $r_i$ and $r_j$. We assume that, for all $i\neq j$, $\mat{SIM}_{i,j} = \mat{SIM}_{j,i} \in [0,a]$\footnote{We will not be concerned with the diagonal values of $\mat{SIM}$, because trivially any record is a `duplicate' of itself, but for definiteness we may assume that, for all $i$, $\mat{SIM}_{i,i}=a$.}. If we use our adjusted (soft) TF-IDF method, $\mat{SIM}$ is given by $\mat{adjST}$ from \eqref{eq:adjST}. In Section~\ref{sec:adjusting} we saw that in that case we even have $\mat{SIM}_{i,j} \in [0,1]$.

Let $\tau \in [0,a]$ be a threshold and let S be the \textit{thresholded similarity score matrix} defined for $i\neq j$ as
\[
S_{i,j} := \begin{cases}  	1,	&\text{if } \mat{SIM}_{i,j} \geq \tau, \\
									0, & \text{if } \mat{SIM}_{i,j} < \tau. \end{cases}
\]
The outcome of our method does not depend on the diagonal values, but for definiteness (and to simplify some computations) we set $S_{i,i}:=1$, for all $i$. If we want to avoid trivial clusterings (i.e. with all records in the same cluster, or with each cluster containing only one record) the threshold value $\tau$ must be chosen in the half-open interval 
\[\big(\underset{i,j: j\neq i}\min\, \mat{SIM}_{i,j},\underset{i,j: j \neq i}\max\, \mat{SIM}_{i,j}\big].\]

If $S_{i,j}=1$, then the records $r_i$ and $r_j$ are clustered together. Note that this is a sufficient, but not necessary condition for two records to be clustered together. For example, if $S_{i,j}=0$, but $S_{i,k} = 1$ and $S_{j,k} = 1$, then $r_i$ and $r_k$ are clustered together, as are $r_j$ and $r_k$, and thus so are $r_i$ and $r_j$. The output of the TGS is a clustering of all the records in the data set, i.e. a collection of clusters, each containing one or more records, such that each record belongs to exactly one cluster.

The choice of $\tau$ is crucial in the formation of clusters. Choosing a threshold that is too low leads to large clusters of records that represent more than one unique entity. Choosing a threshold that is too high breaks the data set into a large number of clusters, where a single entity may be represented by more than one cluster. Here, we propose a method of choosing $\tau$.

Let $H \in \R^n$ be the $n\times1$ vector defined by
\begin{equation*}
H_i := \max_{\substack{1\leq j \leq n\\ j\neq i}}\, \mat{SIM}_{i,j}.
\end{equation*}
In other words, the $i^{\text{th}}$ element of $H$ is the maximum similarity score $\mat{SIM}_{i,j}$ between the $i^{\text{th}}$ record and every other record. Now define
\begin{equation*}
\tau_H := \begin{cases}  	
\mu(H) + \sigma(H), &\hspace{-0.27cm}\text{if } \mu(H) +\sigma(H) < \max_i H_i, \\
\mu(H),	&\hspace{-0.27cm}\text{else},
\end{cases}
\end{equation*}
where $\mu(H)$ is the mean value of $H$ and $\sigma(H)$ is its corrected sample standard deviation\footnote{We used MATLAB's \texttt{std} function.}.

We choose $\tau_H$ in this fashion, because it is easily implementable, has shown to work well in practice (see Section~\ref{sec:Applications}) even if it is not always the optimal choice, and is based on some underlying heuristic ideas and empirical observations of the statistics of $H$ in our data sets (which we suspect to be more generally applicable to other data sets) that we will explain below. It provides a good alternative to trial-and-error attempts at finding the optimal $\tau$, which can be quite time-intensive.

\begin{figure}
 \centering
 \begin{subfigure}[b]{0.45\textwidth}
 \includegraphics[width=\textwidth]{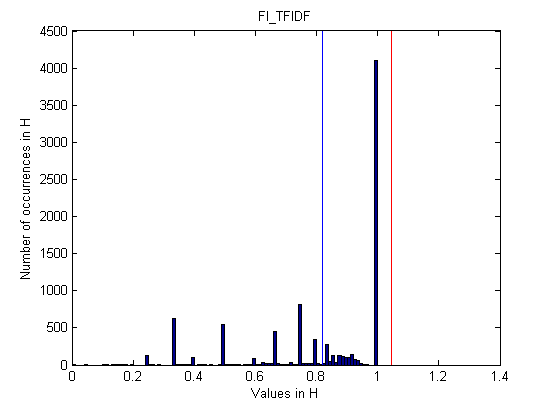}
\caption{$H$ corresponding to the TF-IDF method (with word feature, without refinement step, see Section~\ref{sec:refinement}) applied to the FI data set. The red line is the chosen value $\tau_H = \mu(H)+\sigma(H)$; the blue line indicates $\mu(H)$. }
\label{fig:resth}
 \end{subfigure}
\quad
\begin{subfigure}[b]{0.45\textwidth}
 \includegraphics[width=\textwidth]{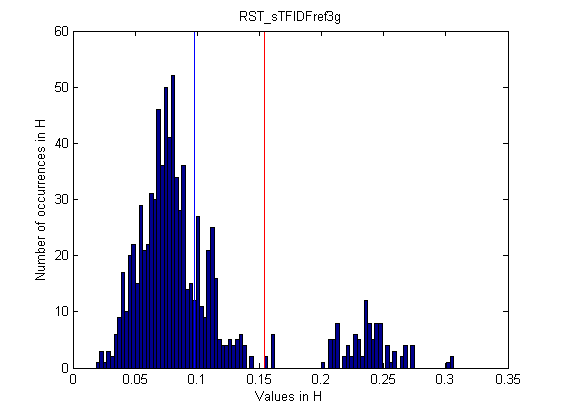}
 \caption{$H$ corresponding to the soft TF-IDF method (with $3$-gram features, with refinement, see Section~\ref{sec:refinement}) applied to the RST data set. The blue line indicates the chosen value $\tau_H = \mu(H)$; the red line indicates $\mu(H)+\sigma(H)$.}
 \label{fig:resth2}
 \end{subfigure}
 \caption{Histograms of $H$ for different methods applied to the FI and RST data sets (see Section~\ref{sec:datasets})}
 \label{fig:restim}
\end{figure}

For a given record $r_i$, the top candidates to be duplicates of $r_i$ are those records $r_j$ for which $\mat{SIM}_{i,j} = H_i$. A typical data set, however, will have many records that do not have duplicates at all. To reflect this, we do not want to set the threshold $\tau_H$ lower than $\mu(H)$. If $H$ is normally distributed, this will guarantee that at least approximately half of the records in the data set will not be clustered together with any other record. In fact, in many of our runs (Fig.~\ref{fig:resth} is a representative example), there is a large peak of $H$ values around the mean value $\mu(H)$. Choosing $\tau_H$ equal to $\mu(H)$ in this case will lead to many of the records corresponding to this peak being clustered together, which is typically not preferred. Choosing $\tau_H = \mu(H)+\sigma(H)$ will place the threshold far enough to the right of this peak to avoid overclustering, yet also far enough removed from the maximum value of $H$ so that not only the top matches get identified as duplicates. In some cases, however, the distribution of $H$ values has a peak near the maximum value instead of near the mean value (as, for example, in Fig.~\ref{fig:resth2}) and the value $\mu(H)+\sigma(H)$ will be larger than the maximum $H$ value. In those cases we can chose $\tau_H = \mu(H)$ without risking overclustering.

\begin{figure}
       \centering
       \begin{subfigure}[b]{0.45\textwidth}
               \includegraphics[width=\textwidth]{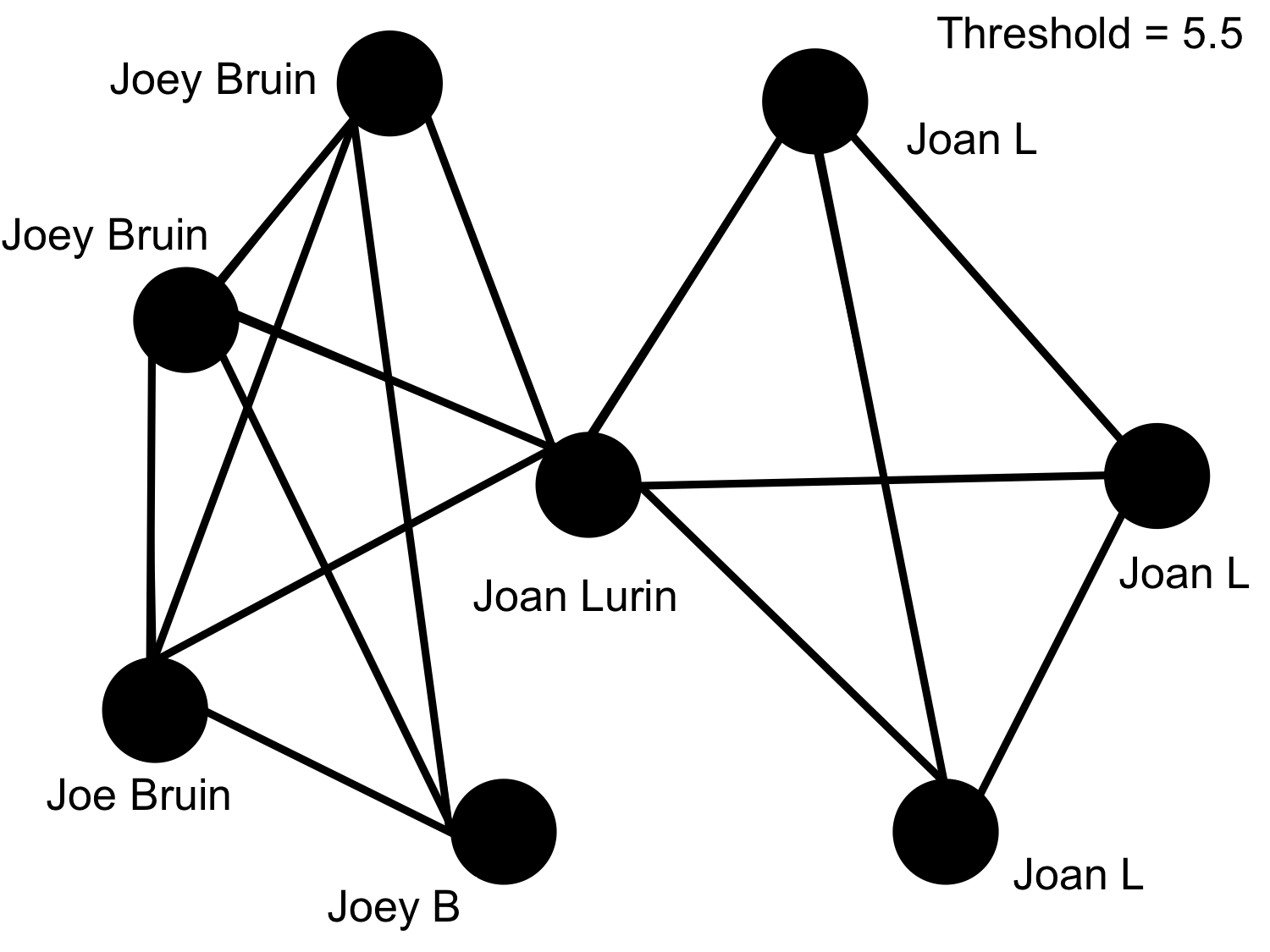}
               \caption{Result of the TGS with $\tau = 5.5$}
               \label{fig:network55}
       \end{subfigure}
       ~ 
       \begin{subfigure}[b]{0.45\textwidth}
               \includegraphics[width=\textwidth]{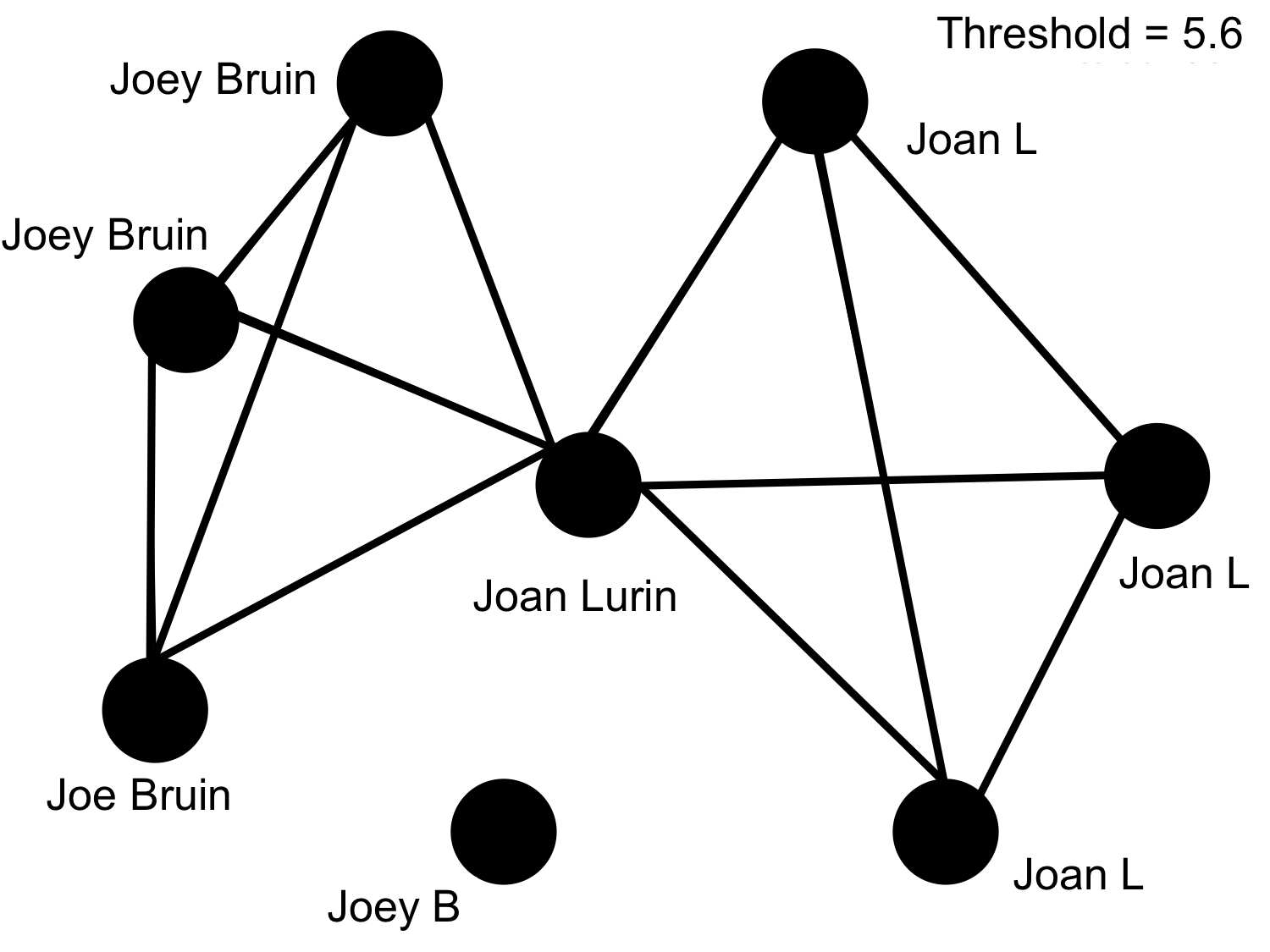}
               \caption{Result of the TGS with $\tau = 5.6$}
               \label{fig:network56}
       \end{subfigure}
       \caption{Two examples of clusters created by the TGS applied to an artificial data set, with different threshold values $\tau$}
       \label{fig:netgraphs}
\end{figure}

It may not always be possible to choose a threshold in such a way that all the clusters generated by our TGS correspond to sets of actual duplicates, as the following example, illustrated in Fig.~\ref{fig:netgraphs}, shows. We consider an artificial toy data set for which we computed the adjusted soft TF-IDF similarity, based on seven fields. We represent the result of the TGS as a graph in which each node represents a record in the data set. We connect nodes $i$ and $j$ ($i\neq j$) by an edge if and only if their similarity score $\mat{SIM}_{i,j}$ equals or exceeds the chosen threshold value $\tau$. The connected components of the resulting graph then correspond to the clusters the TGS outputs.

For simplicity, Fig.~\ref{fig:netgraphs} only shows the features of each entry from the first two fields (first name and last name). Based on manual inspection, we declare the ground truth for this example to contain two unique entities: ``Joey Bruin" and ``Joan Lurin". The goal of our TGS is to detect two clusters, one for each unique entity. Using $\tau = 5.5$, we find one cluster (Fig.~\ref{fig:network55}). Using $\tau = 5.6$, we do obtain two clusters (Fig.~\ref{fig:network56}), but it is not true that one cluster represents ``Joey Bruin'' and the other ``Joan Lurin'', as desired. Instead, one clusters consists of only the ``Joey B'' record, while the other cluster contains all other records. Increasing $\tau$ further until the clusters change, would only result in more clusters, therefore we cannot obtain the desired result this way. This happens because the adjusted soft TF-IDF similarity between ``Joey B" and ``Joey Bruin" (respectively ``Joe Bruin'') is less than the adjusted soft TF-IDF similarity between ``Joey Bruin" (respectively ``Joe Bruin'') and ``Joan Lurin". To address this issue, we apply a \textit{refinement step} to each set of clustered records created by the TGS, as explained in the next section.

The graph representation of the TGS output turns out to be a very useful tool and we will use its language in what follows interchangeably with the cluster language.

\vspace{.5cm}
\begin{algorithm}
\KwData{$\mat{SIM} = \mat{ST} \in \mathbb{R}^{n \times n}$, threshold value $\tau$ (manual choice or automatic $\tau=\tau_H$)}
\KwResult{a collection of $c$ clusters $\mathcal{C}=\{R_1 \ldots R_c\}$}
\For{each $i$}{
    \emph{Initialize} $S_{i,i}=1$}
\For{each pair of distinct records $r_i$ and $r_j$}{
	\emph{Compute $S_{i,j}$}\\
}
\For{each pair of distinct records $r_i$ and $r_j$}{
    If $S_{i,j} = 1$, assign $r_i$ and $r_j$ to the same cluster}
\caption{Thresholding and grouping}
\end{algorithm}

\subsection{Refinement}\label{sec:refinement} As the discussion of the TGS and the example in Fig.~\ref{fig:netgraphs} have shown, the clusters created by the TGS are not necessarily complete subgraphs: it is possible for a cluster to contain records $r_i$, $r_j$ for which $S_{i, j} = 0$.  In such cases it is a priori unclear if the best clustering is indeed achieved by grouping $r_i$ and $r_j$ together or not. We introduce a way to refine clusters created in the TGS, to deal with situations like these. We take the following steps to refine a cluster $R$:
\begin{enumerate}
\item determine whether $R$ needs to be refined by determining the cluster stability with respect to single record removal;
\item if $R$ needs be to refined, remove one record at a time from $R$ to determine the `optimal record' $r^*$ to remove;
\item if $r^*$ is removed from $R$, find the subcluster that $r^*$ does belong to.
\end{enumerate}
Before we describe these steps in more detail, we introduce more notation. Given a cluster (as determined by the TGS) $R = \{ r_{t_1}, \ldots, r_{t_p}\}$ containing $p$ records, the thresholded similarity score matrix of the cluster $R$ is given by the restricted matrix $S|_R \in \mathbb{R}^{p \times p}$ with elements $(S|_R)_{i,j} := S_{t_i, t_j}$. Remember we represent $R$ by a graph, where each node corresponds to a record $r_{t_i}$ and two distinct nodes are connected by an edge if and only if their corresponding thresholded similarity score $(S|_R)_{i,j}$ is 1. If a record $r_{t_i}$ is removed from $R$, the remaining set of records is\\ $R(r_{t_i}) := \{r_{t_1}, \ldots, r_{t_{i-1}}, r_{t_{i+1}}, \ldots, r_{t_p}\}$. We define the \textit{subclusters} $R_1, \ldots R_q$ of $R(r_{t_i})$ as the subsets of nodes corresponding to the connected components of the subgraph induced by $R(r(t_i))$.

\begin{figure*}[htp]
\centering
\includegraphics[scale=.35]{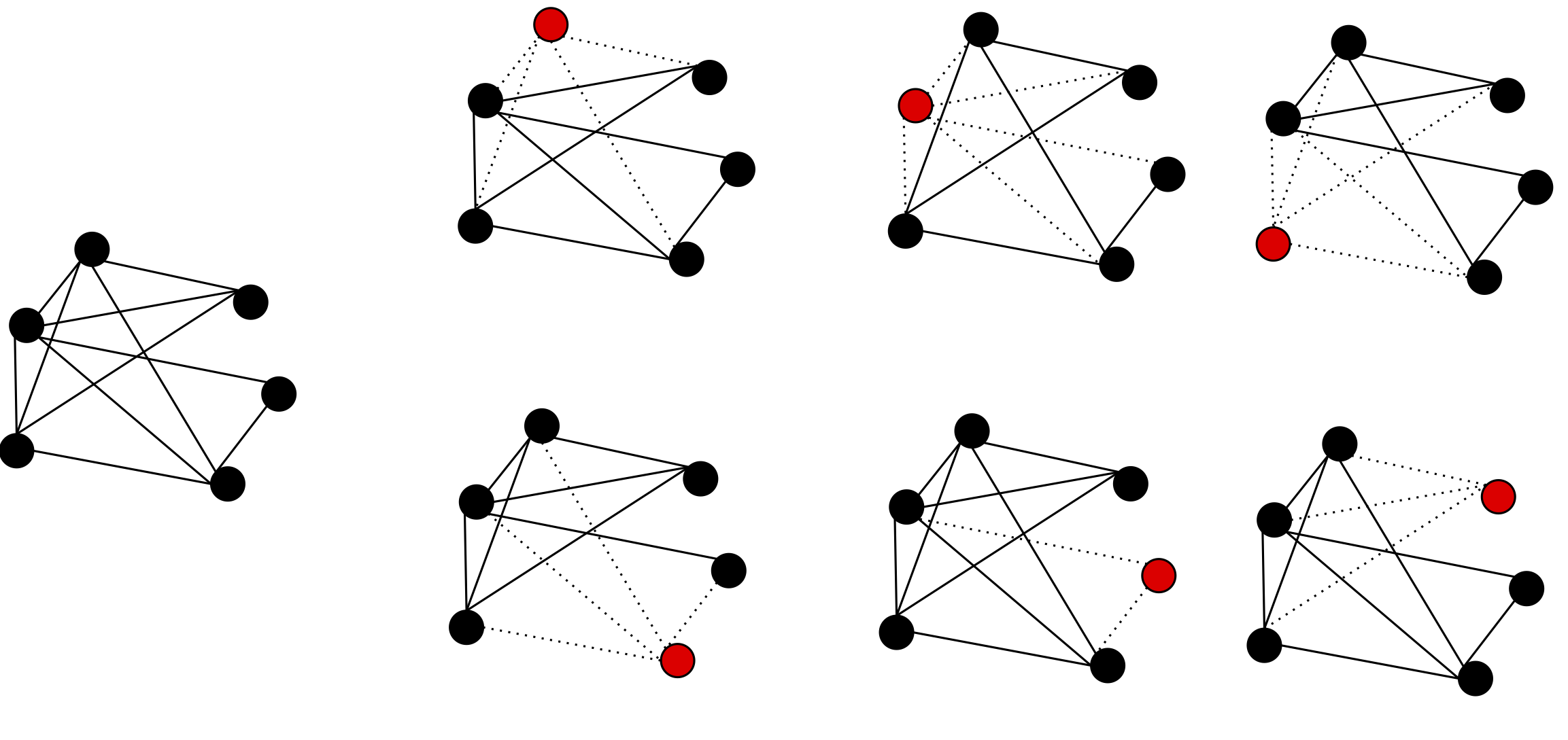}
\caption{An example of a cluster $R$ that does not require refinement. Each node represents a record. In each test we remove one and only one node from the cluster and apply TGS again. The red node represents the removed record $r_{t_i}$, the remaining black nodes make up the set $R(t_i)$.  Notice that every time we remove a record, all other records are still connected to each other by solid lines, hence $R$ does not need to be refined.}
\label{fig:network-nr}
\end{figure*}

\paragraph{Step 1.} Starting with a cluster $R$ from the TGS, we first determine if $R$ needs to be refined, by investigating, for each $r_{t_i} \in R$, the subclusters of $R(r_{t_i})$. If, for every $r_{t_i} \in R$, $R(r_{t_i})$ has a single subcluster, then $R$ need not be refined. An example of this is shown in Fig.~\ref{fig:network-nr}. If there is an $r_{t_i}\in R$, such that $R(r_{t_i})$ has two or more subclusters, then we refine $R$.

\paragraph{Step 2.} For any set $\tilde R$ consisting of $p$ records, we define its \textit{strength} as the average similarity between the records in $\tilde R$:
\begin{equation}\label{eq:strength}
s(\tilde R) := \begin{cases}
\frac{\sum\limits_{i,j=1 \atop i \not= j}^p (S|_{\tilde R})_{i,j}}{{p \choose 2}}, & \text{if } p\geq 2,\\
0, & \text{if } p=1.
\end{cases}
\end{equation}
Note that $s(\tilde R)$ = 1 if $S|_{\tilde R} = \textbf{1}^{p\times p}$ (it suffices if the off-diagonal elements satisfy this equality). In other words, a cluster has a strength of 1 if every pair of distinct records in that cluster satisfy condition 1 of the TGS.

If in Step 1 we have determined that the cluster $R$ requires refinement, we find the optimal record $r^* := r_{t_{k^*}}$ such that the average strength of subclusters of $R(r^*)$ is maximized:
\begin{equation*}
 k^* = \argmax_{1 \leq i \leq p} \frac{1}{q(i)}\sum_{j=1}^{q(i)} s(R_j).
\end{equation*}
Here the sum is over all $j$ such that $R_j$ is a subcluster of $R(r_{t_i})$, and $q(i)$ is the ($i$-dependent) number of subclusters of $R(r_{t_i})$. In the unlikely event that the maximizer is not unique, we arbitrarily choose one of the maximizers as $k^*$. Since the strength of a subcluster measures the average similarity between the records in that subcluster, we want to keep the strength of the remaining subclusters as high as possible after removing $r^*$ and optimizing the average strength is a good strategy to achieve that. 

\paragraph{Step 3.} After finding the optimal $r^*$ to remove, we now must determine the subcluster to which to add it. We again use the strength of the resulting subclusters as a measure to decide this. We evaluate the strength of the set $R_j \cup \{r^*\} \subset R$, for each subcluster $R_j \subset R(r^*)$. We then add $r^*$ to subcluster $R_{l^*}$ to form $R^*:=R_{l^*} \cup \{r^*\}$, where
\[
l^* := \argmax_{j: \, R_j \text{ is a subcluster} \atop \text{ of } R(r^*)} s(R_j \cup \{r^*\}).
\]
In the rare event that the maximizer is not unique, we arbitrarily choose one of the maximizers as $l^*$. Choosing $l^*$ in this way ensures that $r^*$ is similar to the records in $R_{l^*}$.

We always add $r^*$ to one of the other subclusters and do not consider the possibility of letting $\{r^*\}$ be its own cluster. Note that this is justified, since from our definition of strength in \eqref{eq:strength}, $s(\{r^*\}) = 0 < s(R^*)$, because $r^*$ was connected to at least one other record in the original cluster $R$.

Finally, the original cluster $R$ is removed from the output clustering, and the new clusters\\ $R_1, \ldots, R_{l^*-1}, R^*, R_{l^*+1}, \ldots, R_{q(k^*)}$ are added to the clustering.

\begin{figure}[htp]
\centering
\includegraphics[scale=.30]{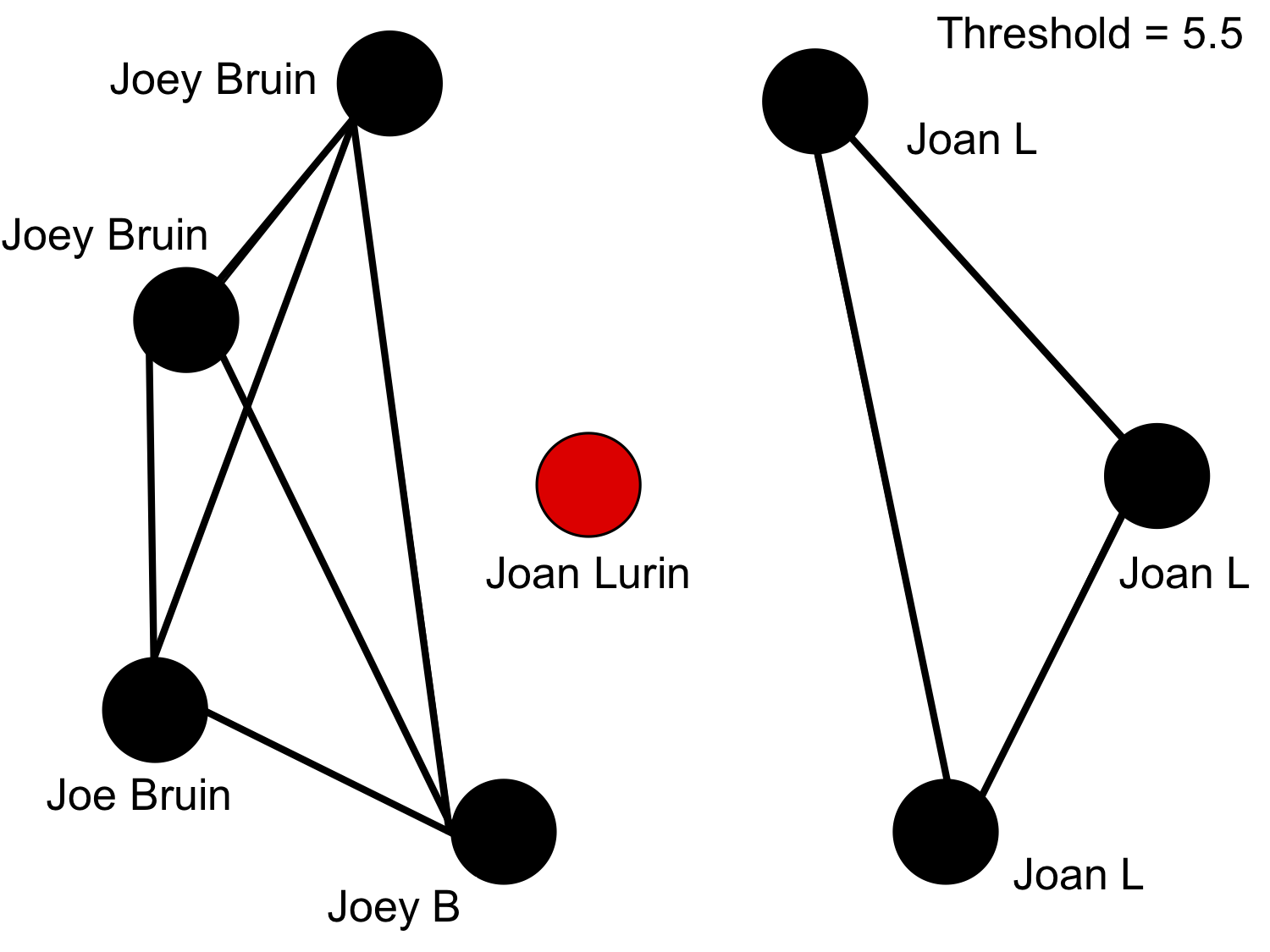}
\quad
\includegraphics[scale=.30]{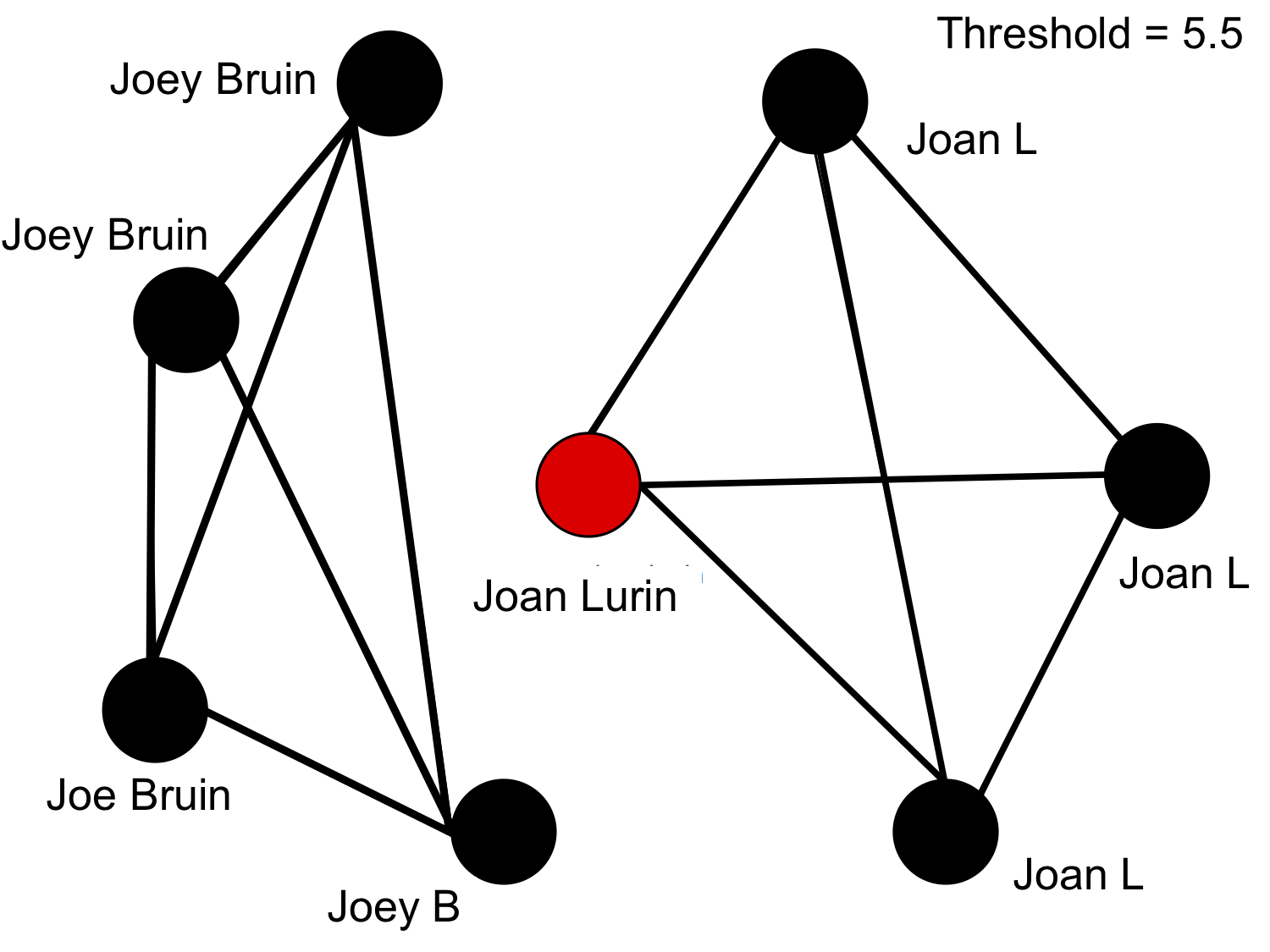}
\caption{An example of how refinement is used to improve our clusters. The left figure shows that by removing the record ``Joan Lurin'', we obtain the two desired subsets. The right figure shows that ``Joan Lurin'' is inserted back into the appropriate cluster. Note that we have not changed the threshold value $\tau$ during this process.}
\label{fig:network-refine}
\end{figure}

Fig.~\ref{fig:network-refine} shows an example of how the refinement helps us to find desired clusters.

\begin{algorithm}
\KwData{$R = \{r_{t_1}, \ldots, r_{t_n}\}$ a cluster resulting from the TGS}
\KwResult{$\mathcal{R}$ set of refined clusters}
\If{there exists $r_{t_i}$ such that $R(r_{t_i})$ has more than 1 subcluster}{
\For{each $r_{t_i} \in R$}{
\emph{Find the subclusters $R_1, \ldots R_q$ of $R(r_{t_i})$}\\
\emph{Compute $\frac{1}{q}\sum_{j=1}^q s(R_j)$}
}
\emph{Assign $r^* = r_{t_{k^*}}$ where $k^* = \argmax_{i} \frac{1}{q}\sum_{j=1}^q s(R_j)$}\\
\For{each subcluster $R_i \subset R(r^*)$}{
\emph{Compute $s(R_i \cup \{r^*\})$}
}
\emph{Assign $R^* = (R_{l^*} \cup \{r^*\})$ where $l^* = \argmax_{j} s(R_j \cup \{r^*\})$}\\
\emph{$\mathcal{R} = \{R_1,\ldots, R_{l^*-1}, R_{l^*}, R_{l^*+1},\ldots, R_{q(k^*)}$\}}
}
\Else{
\emph{Do not refine R: $\mathcal{R} = \{R\}$}
}
\caption{Refinement}
\end{algorithm}

In our implementation, we computed the optimal values $k^*$ and $l^*$ are via an exhaustive search over all parameters. This can be computationally expensive when the initial threshold $\tau$ is small, leading to large initial clusters.

We only applied the refinement step process once (i.e., we executed Step 1 once and for each cluster identified in that step we applied Steps 2 and 3 once each). It is possible to iterate this three step process until no more `unstable' clusters are found in Step 1.


\section{Results}\label{sec:Applications}

\subsection{The data sets}\label{sec:datasets}
The results presented in this section are based on four data sets: the \textit{Field Interview Card data set} (FI), the \textit{Restaurant data set} (RST), the \textit{Restaurant data set with entries removed to induce sparsity} (RST30), and the \textit{Cora Citation Matching data set} (Cora). FI is not publicly available. The other data sets currently can be found at \cite{datasets}. Cora can also be accessed at \cite{cora}. RST and Cora are also used in \cite{bilenko2003adaptive} to compare several approaches to evaluate duplicate detection.

\paragraph{FI} This data set consists of digitized \textit{Field Interview cards} from the LAPD. Such cards are created at the officer's discretion whenever an interaction occurs with a civilian. They are not restricted to criminal events. Each card contains 61 fields of which we use seven: last name, first name, middle name, alias/moniker, operator licence number (driver's licence), social security number, and date of birth. A subset of this data set is used and described in more detail in \cite{vanGennipHunteretal13}. The FI data set has 8,834 records, collected during the years 2001--2011. A ground truth of unique individuals is available, based on expert opinion. There are 2,920 unique people represented in the FI data set. The FI data set has many misspellings as well as different names that correspond to the same individual. 
Approximately 30$\%$ of the entries are missing, but the ``last name'' field is without missing entries. 

\paragraph{RST} This data set is a collection of restaurant information based on reviews from \textit{Fodor} and \textit{Zagat}, collected by Dr. Sheila Tejada \cite{tejada01}, who also manually generated the ground truth. It contains five fields: restaurant name, address, location, phone number, and type of food. There are 864 records containing 752 unique entities/restaurants. There are no missing entries in this data set. The types of errors that are present include word and letter transpositions, varying standards for word abbreviation (e.g. ``deli'' and ``delicatessen''), typographical errors, and conflicting information (such as different phone numbers for the same restaurant). 

\paragraph{RST30}To be able to study the influence of sparsity of the data set on our results, we remove approximately 30\% of the entries from the address, city, phone number, and type of cuisine fields in the RST data set. The resulting data set we call RST30. We choose the percentage of removed entries to correspond to the percentage of missing entries in the FI data set. Because the FI data set has a field that has no missing entries, we do not remove entries from the ``name'' field. 

\paragraph{Cora} The records in the \textit{Cora Citation Matching data set}\footnote{The Cora data set should not be confused with the \textit{Coriolis Ocean database ReAnalysis} (CORA) data set.} are citations to research papers \cite{mccallum2000}. Each of Cora's 1,295 records is a distinct citation to any one of the 122 unique papers to which the data set contains references. We use three fields: author(s), name of publication, and venue (name of the journal in which the paper is published). This data set contains misspellings and a small amount of missing entries (approximately 3\%).

\subsection{Evaluation metrics}

We compare the performances of the methods summarized in Table \ref{tab:methods}. Each of these method outputs a similarity matrix, which we then use in the TGS to create clusters.

To evaluate the methods, we use \textit{purity} \cite{harris1999}, \textit{inverse purity}, their \textit{harmonic mean} \cite{gonzalez2008}, the \textit{relative error in the number of clusters}, \textit{precision}, \textit{recall} \cite{cohen2002learning,bilenko2003evaluation}, the \textit{F-measure} (or \textit{F$_1$ score}) \cite{vanrijsbergen1979,baeza1999modern}, \textit{$z$-Rand score} \cite{meila2007,traud2011comparing}, and \textit{normalized mutual information} (NMI) \cite{strehl2002}, which are all metrics that compare the output clusterings of the methods with the ground truth.

Purity and inverse purity compare the clusters of records which the algorithm at hand gives with the ground truth clusters. Let $\mathcal{C} := \{R_1, \ldots, R_c\}$ be the collection of $c$ clusters obtained from a clustering algorithm and let $\mathcal{C'} := \{R_1', \ldots, R_{c'}'\}$ be the collection of $c'$ clusters in the ground truth. Remember that $n$ is the number of records in the data set. Then we define \textit{purity} as
\[
\mat{Pur}(\mathcal{C}, \mathcal{C}') := \frac1n \sum_{i=1}^c \underset{1\leq j \leq c'}\max\, |R_i \cap R_j'|,
\]
where we use the notation $|A|$ to denote the cardinality of a set $A$. In other words, we identify each cluster $R_i$ with (one of the) ground truth cluster(s) $R_j'$ which shares the most records with it, and compute purity as the total fraction of records that is correctly classified in this way. Note that this measure is biased to favor many small clusters over a few large ones. In particular, if each record forms its own cluster, $\mat{Pur}=1$. To counteract this bias, we also consider \textit{inverse purity},
\[
\mat{Inv}(\mathcal{C}, \mathcal{C}') := \mat{Pur}(\mathcal{C}', \mathcal{C}) = \frac1n \sum_{i=1}^{c'} \underset{1\leq j \leq c}\max\, |R_i' \cap R_j|.
\]
Note that inverse purity has a bias that is opposite to purity's bias: if the algorithm outputs only one cluster containing all the records, then $\mat{Inv}=1$.

We combine purity and inverse purity in their \textit{harmonic mean}\footnote{The harmonic mean of purity and inverse purity is sometimes also called the F-score or $F_1$-score, but we will refrain from using this terminology to not create confusion with the harmonic mean of precision and recall.},
\[
\mat{HM}(\mathcal{C}, \mathcal{C}') := \frac{2 \mat{Pur} \times \mat{Inv}}{\mat{Pur} + \mat{Inv}}.
\]

The \text{relative error in the number of clusters} in $\mathcal{C}$ is defined as
\[
\frac{\big| |\mathcal{C}|-|\mathcal{C}'| \big|}{|\mathcal{C}'|} = \frac{|c-c'|}{c'}.
\]

We define precision, recall, and the F-measure (or F$_1$ score) by considering \textit{pairs} of clusters that have correctly been identified as duplicates. This differs from purity and inverse purity as defined above, which consider individual records. To define these metrics the following notation is useful. Let $G$ be the set of (unordered) pairs of records that are duplicates, according to the ground truth of the particular data set under consideration,
\[
G := \big\{ \{r, s\}: r\neq s \text{ and } \exists R' \in \mathcal{C}' \text{ s. t. } r, s \in R'\},
\]
and let $C$ be the set of (unordered) record pairs that have been clustered together by the duplicate detection method of choice,
\[
C := \big\{ \{r, s\}: r\neq s \text{ and } \exists R \in \mathcal{C} \text{ s. t. } r, s \in R\big\}.
\]

\textit{Precision} is the fraction of the record pairs that have been clustered together that are indeed duplicates in the ground truth,
\[
\mat{Pre}(\mathcal{C}, \mathcal{C}') := \frac{|C \cap G|}{|C|},
\]
and \textit{recall} is the fraction of record pairs that are duplicates in the ground truth that have been correctly identified as such by the method
\[
\mat{Rec}(\mathcal{C}, \mathcal{C}') := \frac{|C \cap G|}{|G|}.
\]
The \textit{F-measure} or \textit{F$_1$ score} is the harmonic mean of precision and recall,
\[
F(\mathcal{C}, \mathcal{C}') := 2 \ \frac{\mat{Pre}(\mathcal{C}, \mathcal{C}') \times \mat{Rec}(\mathcal{C}, \mathcal{C}')}{\mat{Pre}(\mathcal{C}, \mathcal{C}') + \mat{Rec}(\mathcal{C}, \mathcal{C}')} = 2 \ \frac{|C \cap G|}{|G| + |C|}.
\]
Note that in the extreme case in which $|\mathcal{C}|=n$, i.e. the case in which each cluster contains only one record, precision, and thus also the F-measure, are undefined.

Another evaluation metric based on pair counting, is the $z$-Rand score. The \textit{$z$-Rand score} $z_R$ is the number of standard deviations by which $|C \cap G|$ is removed from its mean value under a hypergeometric distribution of equally likely assignments with the same number and sizes of clusters. For further details about the $z$-Rand score, see \cite{meila2007,traud2011comparing,vanGennipHunteretal13}. The \textit{relative $z$-Rand score} of $\mathcal{C}$ is the $z$-Rand score of that clustering divided by the $z$-Rand score of $\mathcal{C}'$, so that the ground truth $\mathcal{C}'$ has a relative $z$-Rand score of 1\footnote{We conjecture that the relative $z$-Rand score is bounded above by 1, but to the best of our knowledge this remains unproven at the moment.}.

A final evaluation metric we consider, is \textit{normalized mutual information} (NMI). To define this, we first need to introduce mutual information and entropy. We define the \textit{entropy} of the collection of clusters $\mathcal{C}$ as
\begin{equation}\label{eq:entropy}
\mat{Ent}(\mathcal{C}) := -\sum_{i=1}^c \frac{|R_i|}n \log\left(\frac{|R_i|}n\right),
\end{equation}
and similarly for $\mat{Ent}(\mathcal{C}')$. The joined entropy of $\mathcal{C}$ and $\mathcal{C}'$ is
\[
\mat{Ent}(\mathcal{C}, \mathcal{C}') := -\sum_{i=1}^c \sum_{j=1}^{c'} \frac{|R_i \cap R_j'|}n \log\left(\frac{|R_i\cap R_j'|}n\right).
\]
The \textit{mutual information} of $\mathcal{C}$ and $\mathcal{C}'$ is then defined as
\begin{align*}
I(\mathcal{C}, \mathcal{C}') &:= \mat{Ent}(\mathcal{C}) + \mat{Ent}(\mathcal{C}') - \mat{Ent}(\mathcal{C}, \mathcal{C}')\\ &= \sum_{i=1}^c \sum_{j=1}^{c'} \frac{|R_i \cap R_j'|}n \log\left(\frac{n |R_i\cap R_j'|}{|R_i| |R_j|}\right),
\end{align*}
where the right hand side follows from the equalities $\sum_{i=1}^c |R_i \cap R_j'| = |R_j'|$ and $\sum_{j=1}^{c'} |R_i \cap R_j'| = |R_i|$.
There are various ways in which mutual information can be normalized. We choose to normalize by the geometric mean of $\mat{Ent}(\mathcal{C})$ and $\mat{Ent}(\mathcal{C}')$ to give the \textit{normalized mutual information}
\[
\mat{NMI}(\mathcal{C}, \mathcal{C}') := \frac{I(\mathcal{C}, \mathcal{C}')}{\sqrt{\mat{Ent}(\mathcal{C})\mat{Ent}(\mathcal{C}')}}.
\]
Note that the entropy of $\mathcal{C}$ is zero, and hence the normalized mutual information is undefined, when $|\mathcal{C}|=1$, i.e. when one cluster contains all the records. In practice this is avoided by adding a small number (e.g. the floating-point relative accuracy \texttt{eps} in MATLAB) to the argument of the logarithm in \eqref{eq:entropy} for $\mat{Ent}(\mathcal{C})$ and $\mat{Ent}(\mathcal{C}')$.

\bigskip

Because we are testing our methods on data sets for which we have ground truth
available, the metrics we use all compare our output with the ground truth. This
would not be an option in a typical application situation in which the
ground truth is not available. If the methods give good results in test cases in which comparison with the ground truth is possible, it increases confidence in
the methods in situations with an unknown ground truth.
Which of the metrics is the most appropriate in any given situation depends
on the needs of the application. For example, in certain situations (for example
when gathering anonymous statistics from a data set) the most important aspect
to get right might be the number of clusters and thus the relative error in the
number of clusters metric would be well suited for use, whereas in other situations
missing out on true positives or including false negatives might carry a high cost,
in which case precision or recall, respectively, or the $F_1$  score are relevant metrics.
For more information on many of these evaluation metrics, see also \cite{amigo2009}.

\subsection{Results}

\begin{table}
\centering
\begin{tabular}{|c|c|c|c|} \hline
	Name & Similarity & Features & Ref. \\
	& matrix & & \\ \hline
	TFIDF & $\mat{ST}$ using \eqref{eq:sTFIDFforTFIDF} & words & no\\ \hline
	TFIDF 3g & $\mat{ST}$ using \eqref {eq:sTFIDFforTFIDF} & 3-grams & no\\ \hline
	sTFIDF & $\mat{ST}$ using \eqref{eq:sTFIDF} & words & no\\ \hline
	sTFIDF 3g & $\mat{ST}$ using \eqref{eq:sTFIDF} & 3-grams & no\\ \hline
	sTFIDF ref & $\mat{ST}$ using \eqref{eq:sTFIDF} & words & yes\\ \hline
	sTFIDF 3g ref & $\mat{ST}$ using \eqref{eq:sTFIDF} & 3-grams & yes\\ \hline
\end{tabular}
\caption{Summary of methods used. The second, third, and fourth columns list for each method which similarity score matrix is used in the TGS, if words or 3-grams are used as features, and if the refinement step is applied after TGS or not, respectively. Equation \eqref{eq:ST} is always used to compute the similarity score, but the important difference is whether the soft TF-IDF matrix from \eqref{eq:sTFIDF} or the TF-IDF matrix from \eqref{eq:sTFIDFforTFIDF} is used in \eqref{eq:ST}. }\label{tab:methods}
\end{table}

In this section we consider six methods: TF-IDF, soft TF-IDF without the refinement step, and soft TF-IDF with the refinement step, with each of these three methods applied to both word features and $3$-gram features. We also consider five evaluation metrics: the harmonic mean of purity and inverse purity, the relative error in the number of clusters, the $F_1$ score, the relative $z$-Rand score, and the NMI. We investigate the results in two different ways: (a) by plotting the scores for a particular evaluation metric versus the threshold values, for the six different methods in one plot and (b) by plotting the evaluation scores obtained with a particular method versus the threshold values, for all five evaluation metrics in one plot. Since this paper does not offer space to present all figures, we show some illustrative plots and describe the main results in the text. In Section~\ref{sec:conclusions} we will discuss conclusions based on these results.

\subsubsection{The methods}\label{sec:themethods}

\begin{figure}
\centering
 \begin{subfigure}[b]{0.45\textwidth}
 \includegraphics[scale=0.4]{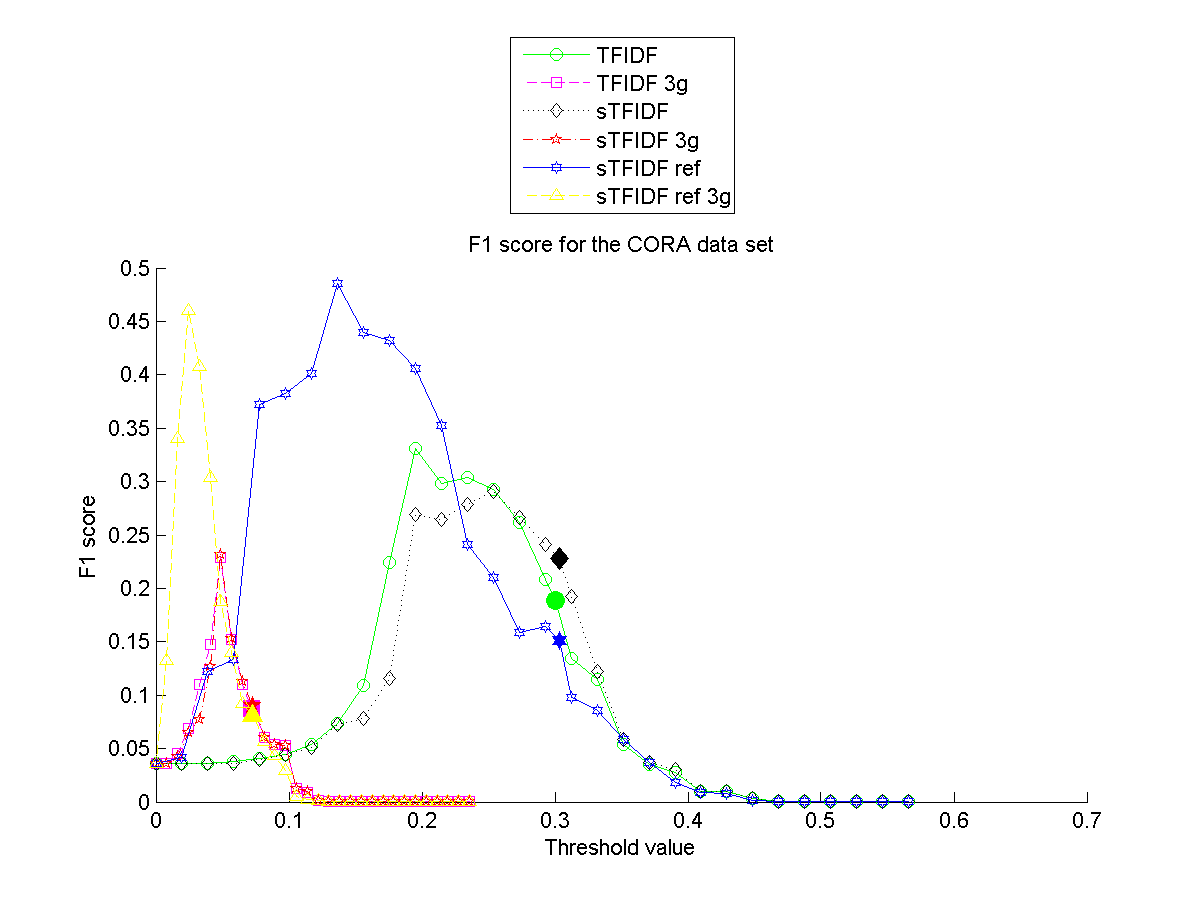}
\caption{The $F_1$ score for the Cora data set\\ \hspace{1cm}}
\label{fig:F1Cora}
 \end{subfigure}
\quad
 \begin{subfigure}[b]{0.45\textwidth}
 \includegraphics[scale=0.5]{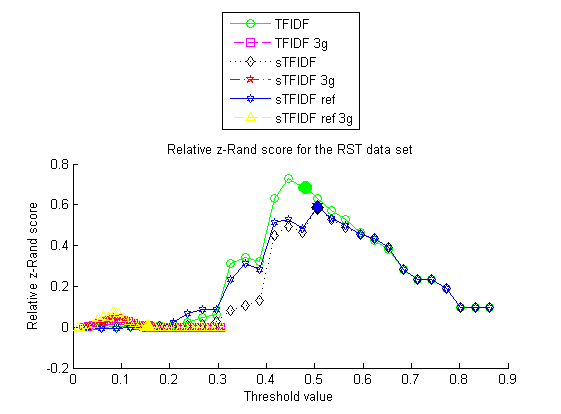}
\caption{The relative $z$-Rand score for the RST data set}\label{fig:zrandRST}
\end{subfigure}
\caption{Two evaluation metrics as a function of the threshold value $\tau$, computed on two different data sets. Each of the six graphs in a plot correspond to one of the six methods used. The filled markers indicate the metric's value at the automatically chosen threshold value $\tau_H$ for each method. In the legend, ``(s)TF-IDF'' stands for (soft) TF-IDF, ``3g'' indicates the use of $3$-gram based features instead of word based ones, and ``ref'' indicates the presence of the refinement step.}
\label{fig:metricplots}
\end{figure}

When we compare the different methods by plotting the scores for a particular evaluation metric versus the threshold value $\tau$ for all the methods in one plot (as can be seen for example in Fig.~\ref{fig:F1Cora}), one notable attribute is that the behavior of the methods that use word features typically is quite distinct from that of the methods that use $3$-gram features. This is not very surprising, since the similarity scores produced by those methods, and hence their response to different threshold values, are significantly different.

It is also interesting to note which methods give better evaluation metric outcomes on which data sets.  First we compare the word based methods with the $3$-gram based methods. On the FI data set the word feature based methods outperform the $3$-gram based methods (judged on the basis of best case performance, i.e. the optimal score attained over the full threshold range) for every evaluation metric by quite a margin, except for the NMI for which the margin is minimal (but still extant).

On both the RST and RST30 data sets, the word feature based methods outperform the $3$-gram feature based methods on the pair counting based metrics, i.e. $F_1$ score and relative $z$-Rand score (Fig.~\ref{fig:zrandRST}), but both groups of methods perform equally well for the other metrics. 

An interesting difference between the Cora data set and the other data sets, is that while sTFIDF ref (see Table~\ref{tab:methods}) does outperform sTFIDF 3g ref on the pair counting based metrics for the Cora data set, the diference is much less pronounced than for the other data sets. The difference in the relative error in the number of clusters is more pronounced however, in favor of the former method. Only on the relative error in the number of clusters does it perform somewhat worse than sTIDF ref. In fact, on all other metrics sTFIDF 3g ref outperforms the other two word based methods (TFIDF and sTFIDF). The other $3$-gram based methods perform worse than their word based counterparts on the pair counting metrics and on par with them on the other metrics. 

Next we compare the TF-IDF methods with the soft TF-IDF methods (without refinement step in all cases). There are very few observable differences between TFIDF 3g and sTFIDF 3g in any of the metrics or data sets, and where there are, the differences are minor. 

The comparison between TFIDF and sTFIDF shows more variable behavior. The most common behavior among all metrics and data sets is that both methods perform equally well in the regions with very small or very large values of $\tau$, although in some cases these regions themselves can be very small indeed. In the intermediate region, TFIDF usually performs better at small $\tau$ values, whereas sTFIDF performs better at larger $\tau$ values. The size of the these different regions, as well as the size of the difference in outcome can differ quite substantially per case. For example, in the case of NMI for the Cora data set, NMI and the harmonic mean of purity and inverse purity for the RST data set, and all metrics except the relative error in the number of clusters for the RST30 data set, TFIDF outperforms sTFIDF quite consistently in the regions where there is a difference.

When it comes to the benefits of including the refinement step, the situation is again somewhat different depending on the data set. First we compare sTFIDF 3g with sTFIDF 3g ref. For small threshold values including the refinement step is beneficial (except in a few cases when there is little difference for very small $\tau$ values). This is to be expected, since the refinement will either increase the number of clusters formed or keep it the same, so its effect is similar to (but not the same as) raising the threshold value. For larger $\tau$ values typically one of two situations occurs: either sTFIDF 3g outperforms sTFIDF 3g ref for intermediate $\tau$ values and there is little difference for higher $\tau$ values, or there is little difference on the whole range of intermediate and large $\tau$ values. The former occurs to a smaller or larger degree for all metrics except NMI for the Cora data set, for the harmonic mean of purity and inverse purity and the relative error in the number of clusters for the FI data set, and also for the relative error in the number of clusters for the RST30 data set. The other cases display the second type of behaviour.

If we compare sTFIDF with sTFIDF ref there are three approximate types of behavior that occur. In the region with very small $\tau$ values the performance is usually similar for both methods, but this region can be very small. Next to this region, there is a region of small $\tau$ values in which sTFIDF ref outperforms sTFIDF. For the same reason as explained above, this is not surprising. This region can be followed by a region of the remaining intermediate and large $\tau$ values in which sTFIDF outperforms sTFIDF ref (the $F_1$ score and harmonic mean of purity and inverse purity for the FI data set), or by a region of the remaining intermediate and large $\tau$ values in which both methods are on par (NMI for the Cora data set, the $F_1$ score, the harmonic mean of purity and inverse purity, and NMI for the RST30 data set, and all metrics for the RST data set), or by first a region of intermediate $\tau$ values on which sTFIDF outperforms sTFIDF ref, followed by a region on which there is little difference between the methods (all other metric/data set combinations).

It is also noteworthy that all methods do significantly worse on RST30 than on RST, when measured according to the pair counting based methods (the $F_1$ and relative $z$-Rand scores), while there is no great difference, if any, measured according to the other metrics. In this context it is interesting to remember that RST30 is created by removing 30\% of the entries from all but one of the fields of RST.

\subsubsection{The metrics}\label{sec:themetrics}

\begin{figure}
\centering
 \begin{subfigure}[b]{0.45\textwidth}
 \includegraphics[scale=0.5]{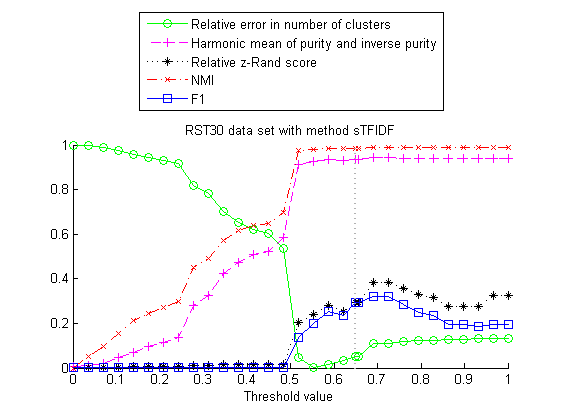}
\caption{Soft TF-IDF (on word based features) without the refinement step applied to the RST30 data set}
\label{fig:sTFIDFRST30}
 \end{subfigure}
\quad
 \begin{subfigure}[b]{0.45\textwidth}
 \includegraphics[scale=0.5]{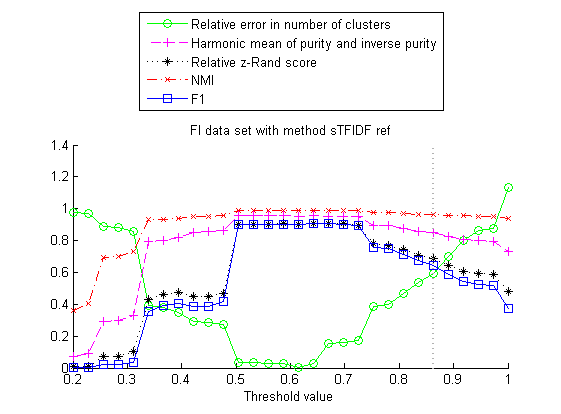}
\caption{Soft TF-IDF (on word based features) with the refinement step applied to the FI data set}
\label{fig:sTFIDFrefFI}
 \end{subfigure}
\caption{Different evaluation metrics as a function of the threshold value $\tau$, computed on two different data sets. Each of the five graphs in a plot correspond to one of five evaluation metrics. The vertical dotted line indicates the automatically chosen threshold value $\tau_H$ for the method used.}
\end{figure}

When plotting the different evaluation metrics per method, we notice that the two pair counting based metrics, i.e. the $F_1$ score and relative $z$-Rand score, behave similarly to each eather, as do the harmonic mean of purity and inverse purity and the NMI. The relative error in the number of clusters is correlated to those other metrics in an interesting way. For the word feature based methods, the lowest relative error in the number of clusters is typically attained at or near the threshold values at which the $F_1$ and relative $z$-Rand scores are highest (this is much less clear for the Cora data set as it is for the others). Those are also usually the lowest threshold values for which the harmonic mean and NMI attain their high(est) values. The harmonic mean and NMI, however, usually remain quite high when the threshold values are increased, whereas the $F_1$ and relative $z$-Rand scores typically drop (sometimes rapidly) at increased threshold values, as the relative error in number of clusters rises. Fig.~\ref{fig:sTFIDFRST30} shows an example of this behavior.

The relationship between the harmonic mean of purity and inverse purity and the NMI has some interesting subtleties. As mentioned before they mostly show similar behavior, but the picture is slightly more subtle in certain situations. On the Cora data set, the harmonic mean drops noticeably for higher threshold values, before settling eventually at a near constant value. This is a drop that is not present in the NMI. This behavior is also present in the plots for the $3$-gram feature based methods on the FI data set and very slightly in the word feature based methods on the RST data set (but not the RST30 data set). For word feature based methods on the FI data set the behavior is even more pronounced, with little to no `settling down at a constant value' happening for high threshold values (e.g. Fig.~\ref{fig:sTFIDFrefFI}).

Interestingly, both the harmonic mean and NMI show very slight (but consistent over both data sets) improvements at the highest threshold values for the $3$-gram based methods applied to the RST and RST30 data sets.

Another meaningful observation is that for $\tau$ values lower than the value at which the relative error in the number of clusters is minimal, TFIDF performs better for this metric than does sTFIDF. This situation is reversed for $\tau$ values higher than the optimal value. This can be understood from the difference between \eqref{eq:sTFIDF} and \eqref{eq:sTFIDFforTFIDF}. Soft TF-IDF incorporates contributions into the similarity score not only from features that are exactly the same in two entries, but also from features that are very similar. Hence the soft TF-IDF similarity score between two entries will be higher than the TF-IDF score between the same entries and thus clusters are less likely to break up at the same $\tau$ value in the soft TF-IDF method than in the TF-IDF method. For $\tau$ values less than the optimal value the breaking up of clusters is beneficial, as the optimal cluster number has not yet been reached and thus TFIDF will outperform sTFIDF on the relative error in the number of clusters metric in this region. For $\tau$ larger than the optimal value, the situation is reversed.

\subsubsection{The choice of threshold}\label{sec:choiceofthresh}

\begin{figure}
\centering
 \begin{subfigure}[b]{0.45\textwidth}
 \includegraphics[scale=0.5]{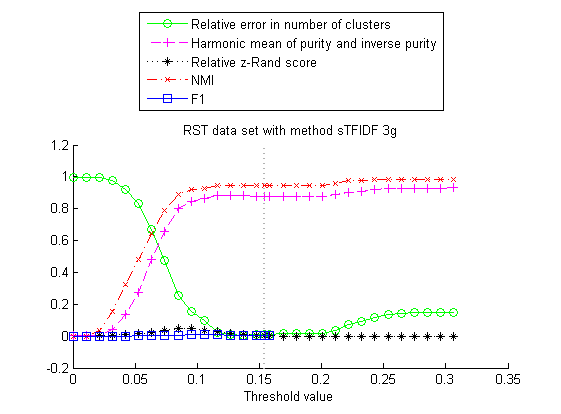}
\caption{Soft TF-IDF (on $3$-gram based features) without the refinement step applied to the RST data set}
\label{fig:sTFIDF3gRST}
 \end{subfigure}
\quad
 \begin{subfigure}[b]{0.45\textwidth}
 \includegraphics[scale=0.5]{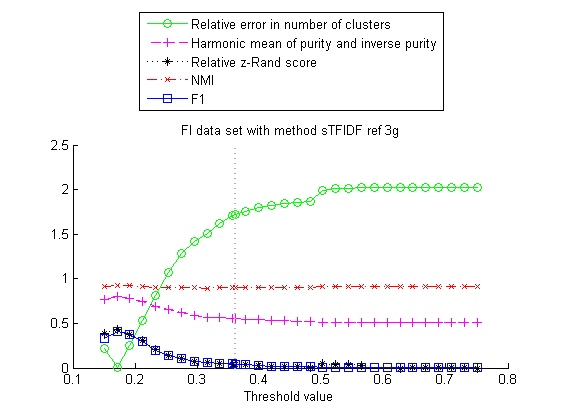}
\caption{Soft TF-IDF (on $3$-gram based features) with the refinement step applied to the FI data set}
\label{fig:sTFIDFref3gFI}
 \end{subfigure}
\caption{Different evaluation metrics as a function of the threshold value $\tau$, computed on two different data sets. Each of the five graphs in a plot correspond to one of five evaluation metrics. The vertical dotted line indicates the automatically chosen threshold value for the method used.}
\end{figure}

On the RST and RST30 data sets our automatically chosen threshold performs well (e.g. see Figs.~\ref{fig:zrandRST}, ~\ref{fig:sTFIDFRST30}, and~\ref{fig:sTFIDF3gRST}). It usually is close to (or sometimes even equal to) the threshold value at which some or all evaluation metrics attain their optimal value (remember this threshold value is not the same for all the metrics). The performance on RST is slightly better then on RST30, as can be expected, but in both cases the results are good.

On the FI and Cora data sets our automatically chosen threshold is consistently larger than the optimal value, as can be seen in e.g. Figs.~\ref{fig:F1Cora}, ~\ref{fig:sTFIDFrefFI}, and~\ref{fig:sTFIDFref3gFI}. This can be explained by the left-skewedness of the $H$-value distribution, as illustrated in Fig.~\ref{fig:resth}. A good proxy for the volume of the tail is the ratio of number of records referring to unique entities to the total number of entries in the data set. For RST and RST30 this ratio is a high 0.87, whereas for FI it is 0.33 and for Cora only 0.09. This means that the relative error in the number of clusters grows rapidly with increasing threshold value and the values of the other evaluation metrics will deteriorate correspondingly.

We also compared whether TFIDF, sTFIDF, or sTFIDF ref performed better at the value $\tau=\tau_H$. Interestingly, sTFIDF ref never outperformed all the other methods. At best it tied with other methods: for the $F_1$ and relative $z$-Rand scores for the RST30 data set it performed equally well as TFIDF; all three methods performed equally well for the NMI for the Cora data set, for the NMI and relative error in the number of clusters for the RST data set, and for NMI and the harmonic mean of the purity and inverse purity for the RST30 data set. TFIDF and sTFIDF tied for the $F_1$ and relative $z$-Rand scores for the FI data set. TFIDF outperformed the other methods on the RST data set for the $F_1$ and relative $z$-Rand scores, as well as the harmonic mean of purity and inverse purity. Finally, sTFIDF outperformed the other methods across the board for the FI data set, as well as for all metrics but the NMI for the Cora data set and for the relative error in the number of clusters for the RST30 data set. To recap, at $\tau=\tau_H$, the soft TF-IDF method seems to be a good choice for the Cora and FI data set, while for most metrics for the RST and RST30 data sets the TF-IDF method is preferred at $\tau=\tau_H$. (Remember that the value $\tau_H$ depends on the data set and the method).

\subsection{Results for alternative sparsity adjustment}\label{sec:imputation}

At the end of Section~\ref{sec:adjusting} we described an alternative sparsity adjustment step, which replaces missing entries by the mode in each field. All the results reported so far use the sparsity adjustment step described in the first part of Section~\ref{sec:adjusting} (which we will call here the ``original'' step); in this section we describe the results obtained using the alternative sparsity adjustment step.

We chose to test this alternative sparstity adjustement step on the Cora and  RST30 data sets. The former has a very small percentage of missing data (approximately $3\%$), while the latter has a high percentage ($30\%$ in all but one of the fields). We use the alternative sparsity adjustment step as part of each of the six methods discussed in this paper. We judge the output again using the same five metrics used above.

\begin{figure}
\centering
 \begin{subfigure}[b]{0.45\textwidth}
 \includegraphics[scale=0.4]{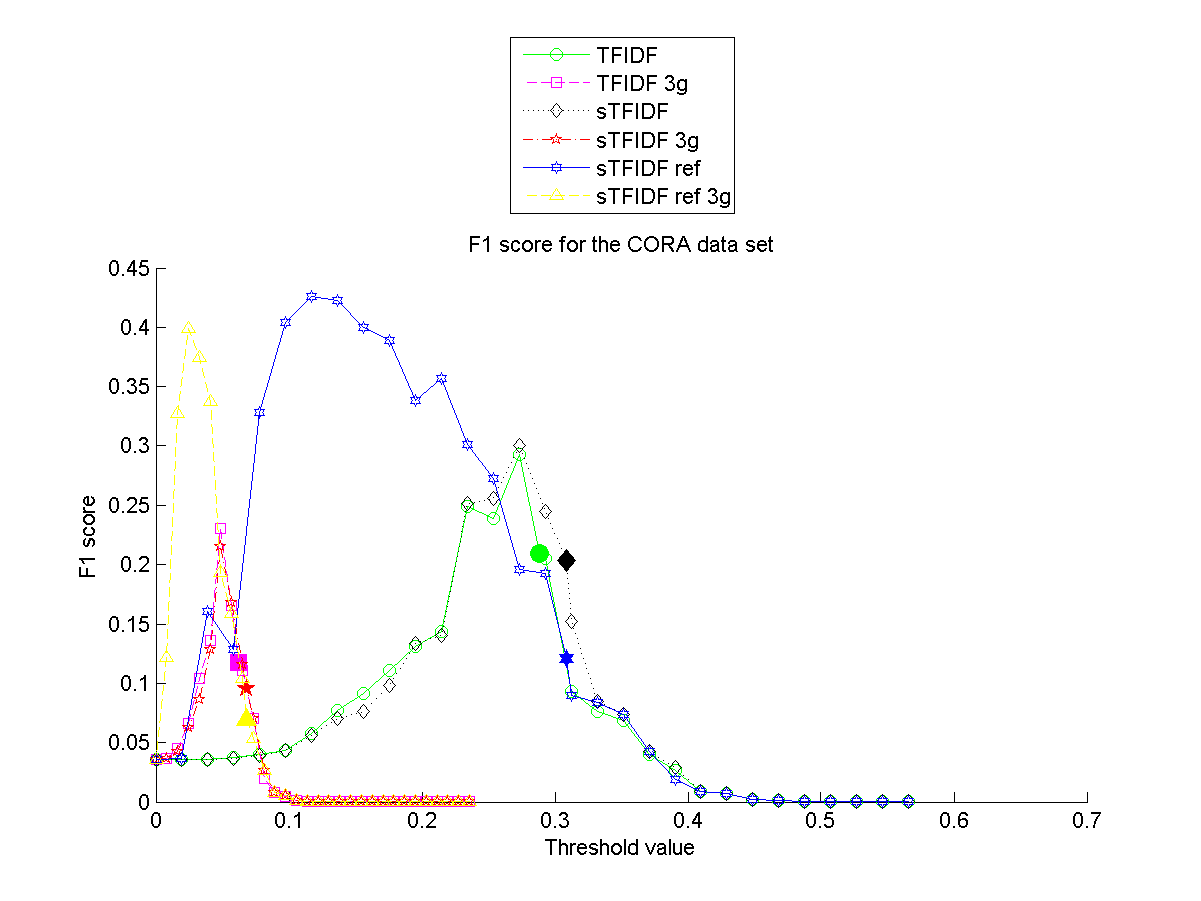}
\caption{The $F_1$ score for the Cora data set; each listed method has the alternative sparsity adjustment step incorporated}
\label{fig:F1Coraalternative}
 \end{subfigure}
 \begin{subfigure}[b]{0.45\textwidth}
 \includegraphics[scale=0.4]{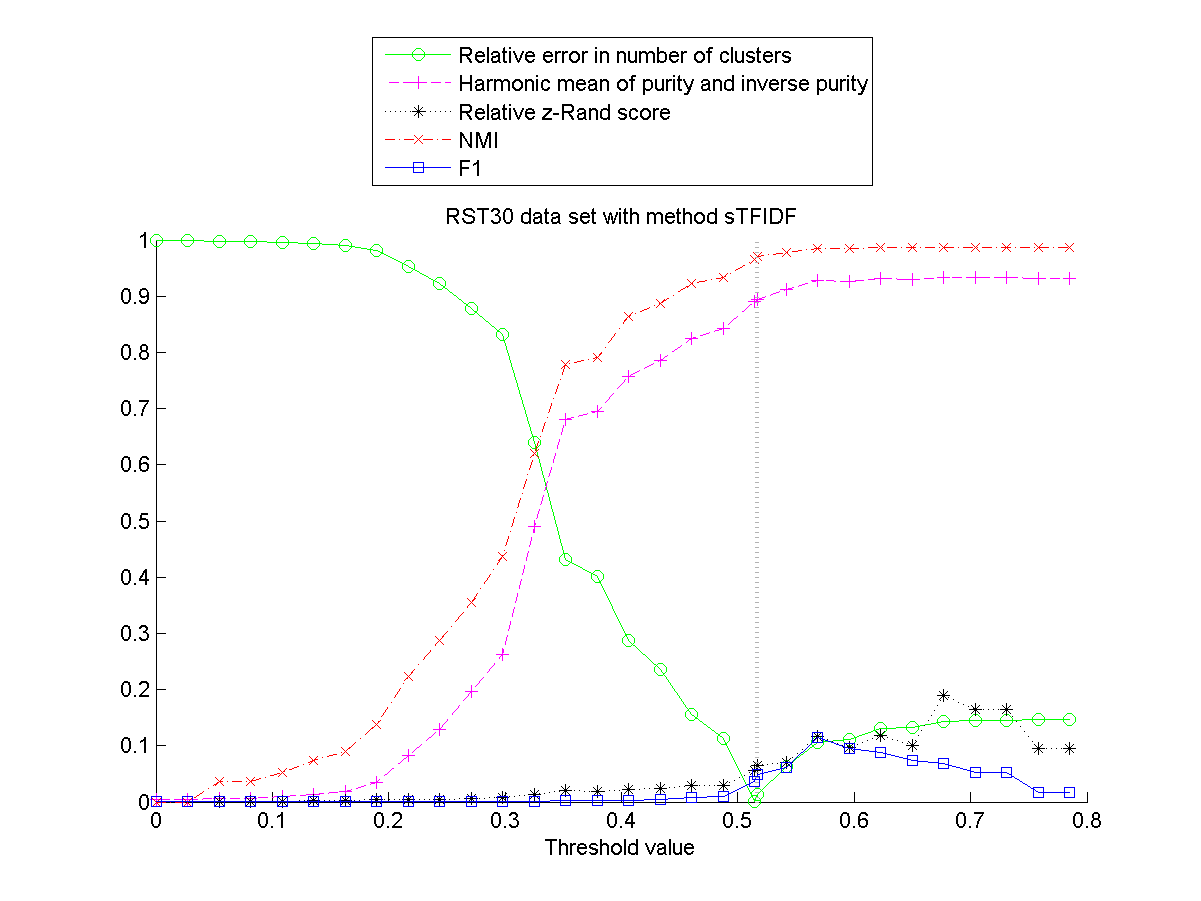}
\caption{Soft TF-IDF (on word based features) without the refinement step applied to the RST30 data set, incorporating the alternative sparsity adjustment step}
\label{fig:sTFIDFRST30alternative}
 \end{subfigure}
\caption{Results obtained using the alternative sparsity adjustment step}
 \end{figure}

In all our tests on the Cora data set there is very little if any difference in the performance of all the methods, with two notable exceptions: the two methods that include the refinement step perform considerably worse according to the two pair counting based metrics (the $F_1$ and relative $z$-Rand scores) when incorporating the alternative sparsity adjustment step (and one minor, yet noticeable exception: TFIDF also performs worse with the alternative adjustment step when measured according to the $F_1$ score). Fig.~\ref{fig:F1Coraalternative} shows the results corresponding to Fig.~\ref{fig:F1Cora}, with as sole difference that in the former the alternative sparsity adjustment step is used, while in the latter the original step is incorporated into the methods.

In all our tests on the RST30 data set the $3$-gram based methods which use the alternative sparsity adjustment step perform very similarly to those that use the original adjustment step (with the difference that those similar results are obtained at lower threshold values when using the alternative step instead of the original adjustement step). The word based methods also perform similarly using either sparsity adjustment step, when measured according to the relative error in the number of clusters, the harmonic mean of purity and inverse purity, and NMI. However, word based methods perform worse with the alternative adjustment step on the pair counting metrics. Fig.~\ref{fig:sTFIDFRST30alternative} shows the results corresponding to the same method as was used in Fig.~\ref{fig:sTFIDFRST30}, with as sole difference the incorporation of the alternative sparsity adjustment step. The worsened performance of the alternative method with respect to the two pair counting metrics can be seen at the high end of the $\tau$-range.

If any general conclusion can be drawn based on these tests, it is that there does not seem to be an advantage in using the alternative sparsity adjustment step instead of the original step; in some cases the resulting output is even worse, when measured according to the pair counting metrics.

A sparsity adjustment method that was not tested in this paper is to replace each missing entry by the same placeholder, e.g. ``[]'' or ``void'' \cite{larose2014discovering}. This in effect will encourage records with missing entries to be clustered together, but carries less risk of them being clustered together with other non-duplicate documents. This could be slightly beneficial in data sets with few missing entries, even though it is effectively a soft version of removing records with missing entries from the data set altogether.

\section{Conclusions and suggestions for future work}\label{sec:conclusions}

In this paper we have investigated six methods which are based on term frequency-inverse document frequency counts for duplicate detection in a record data set. We have tested them on four different data sets and evaluated the outcomes using five different metrics.

One conclusion from our tests is that there is no clear benefit to constructing the features the methods work on using $3$-grams as opposed to white space separated `words'. Keeping the other choices (TF-IDF or soft TFIDF, refinement step or not) the same, using words for the features either outperforms the corresponding $3$-gram based method or performs equally well at worst (in terms of the optimal values that are achieved for the evaluation metrics). See, for example, the graphs in Fig.~\ref{fig:metricplots} or compare Figs.~\ref{fig:sTFIDFrefFI} and~\ref{fig:sTFIDFref3gFI}.

Somewhat surprisingly, our tests lead to a less clear picture regarding the choice between TF-IDF and soft TF-IDF (with word based features, without the refinement step). For low to intermediate threshold values TF-IDF performs better, for higher threshold values either soft TF-IDF performs better, or the difference between the two methods is so small as to be negligible. This behavior is not always very pronounced and, as described in Section~\ref{sec:themethods}, there are even cases in which TF-IDF outperforms soft TF-IDF for almost every threshold value.

The question whether or not to include the refinement step into a (word based) soft TF-IDF method also requires some care. At low $\tau$ values inclusion of the refinement step is beneficial, but at higher values the behavior can vary substantially per data set and metric, as described in Section~\ref{sec:themethods}. As a rule of thumb (but not a hard and fast rule) we can say that for the Cora and FI data sets there is a region of intermediate and/or high $\tau$ values at which including the refinement step is detrimental, whereas for the RST and RST30 data sets soft TF-IDF with refinement at worst performs similar to soft TF-IDF without refinement, but it performs better for certain $\tau$ values as well. This might partly be explained by the observation made in Section~\ref{sec:choiceofthresh}: the FI and Cora data sets have a much lower ratio of unique entities to total number of entries than the RST and RST30 data sets have. Since the refinement step creates extra clusters, including it can be detrimental for data sets that are expected to contain relatively few unique entries. This suspicion is strengthened by the fact that we see in our experiments that the growth in the relative error of the number of clusters when $\tau$ is increased past its optimal value (for that metric) is much larger for the FI and Cora data sets than for the RST and RST30 data sets.

Our tests with our automatically chosen threshold show that $\tau_H = \mu(H) + \sigma(H)$  is a good choice on data sets which have $H$-distributions that are approximately normal or right-skewed. If, however, the $H$-distribution is left-skewed, this choice seems to be consistently larger than the optimal threshold. It should be noted though that for most of the evaluation metrics and most of the data sets, the behavior of the metrics with respect to variations in the threshold value is not symmetric around the optimal value. Typically the decline from optimality is less steep and/or smaller for higher threshold values than for lower ones. This effect is even stronger if we consider methods without refinement step. Combined with the fact that at low threshold values the refinement step requires a lot more computational time than at high threshold values, especially for larger data sets, we conclude that, in the absence of a priori knowledge of the optimal threshold value, it is better to overestimate than underestimate this value. Hence, our suggestion to choose $\tau_H=\mu(U)+\sigma(H)$ is a good rule of thumb at worst and a very good choice for certain data sets.

Since our automated threshold value $\tau_H$ is usally a value in the intermediate or higher end of the $\tau$ range, the discussion above suggests that at $\tau=\tau_H$ it is typically beneficial to use either TF-IDF or soft TF-IDF, in either case without the refinement step. The former is preferred for data sets with a high ratio of unique entities to number of entries, whereas the latter is preferred when this ratio is low. This is consistent with the observations at the end of Sections~\ref{sec:themetrics} (since $\tau_H$ is close to the optimal $\tau$ value where the number of clusters is concerned for the RST and RST30 data sets and overshoots the optimal value for the Cora and FI data sets) and~\ref{sec:choiceofthresh}. This should only be treated as guidance and not as a hard and fast rule.

Future work could explore the possibilities of using methods that first project the data into a lower dimensional latent variable space to allow for duplicate detection in very high dimensional and large data sets, e.g. topic modelling techniques such as latent Dirichlet allocation \cite{BleiNgJordan2003} and nonnegative matrix factorization \cite{KimPark2008}, or the CenKNN method from \cite{PangJinJiang2015}. An overview of other such methods is given in \cite{Dumais2004}. Where possible, new scalable hashing methods that allow for approximate matching might also be considered to reduce computational complexity in such settings \cite{ChiZhu2017}. These methods could reduce the number of comparisons made by quickly identifying specific subsets of pairs (e.g. those that must have similarity zero), but the construction of efficient hash functions is non-trivial and usually domain dependent. Further, the hash functions themselves incur a computational cost, so there is no guarantee of an overall speed up. Finding the right hash function for a given application and exploring the potential benefits of its use in a preprocessing step can be a topic for future research.

\paragraph{Acknowledgements} 
We would very much like to thank George E. Tita and Matthew A. Valasik for their involvement in the collection of the FI data set and the construction of a ground truth clustering. We are grateful to them, as well as to P. Jeffrey Brantingham, for many fruitful discussions. We would also like to thank Brendan Schneiderman, Cristina Garcia-Cardona, and Huiyi Hu for their participation in the 2012 summer Research Experience for Undergraduates (REU) project from which this paper grew.

We would also like to thank the two anonymous reviewers for their very insightful feedback which has lead to significant improvements in the published version of this paper. In particular, we thank one of the reviewers of the first draft of our paper for the suggestion to compare our method with the method of replacing missing entries in a field with the mode of that field. 

This research is made possible via ONR grant N00014-16-1-2119, City of Los Angeles, Gang Reduction Youth Development (GRYD) Analysis Program, AFOSR MURI grant FA9550-10-1-0569, and NSF grants DMS-1045536, DMS-1417674, and DMS-1737770.

Additionally, the second author received  research grants from Claremont McKenna College, the Claremont University Consortium, Alfred P. Sloan Foundation and the NSF, which were not directly related to this research. 

We would like to acknowledge the following sources which were incorporated into or adapted for use in our code: \cite{chen2010nmi,fiedler2010cell,koehler2004matrix,komarov2010set} and \cite{traud2011comparing,traud2011zrand}.

\newpage

\bibliographystyle{spmpsci}

\begin{thebibliography}{10}
\providecommand{\url}[1]{{#1}}
\providecommand{\urlprefix}{URL }
\expandafter\ifx\csname urlstyle\endcsname\relax
  \providecommand{\doi}[1]{DOI~\discretionary{}{}{}#1}\else
  \providecommand{\doi}{DOI~\discretionary{}{}{}\begingroup
  \urlstyle{rm}\Url}\fi

\bibitem{AbrahamKanmani2014}
Abraham, A.A., Kanmani, S.D.: A survey on various methods used for detecting
  duplicates in {XML} data.
\newblock International Journal of Engineering Research \& Technology
  \textbf{3}(1), 354--357 (2014)

\bibitem{ahmed2010dynamic}
Ahmed, I., Aziz, A.: Dynamic approach for data scrubbing process.
\newblock International Journal on Computer Science and Engineering
  \textbf{2}(02), 416--423 (2010)

\bibitem{allison2005imputation}
Allison, P.D.: Imputation of categorical variables with {PROC MI}.
\newblock In: SUGI 30 Proceedings. SAS Institute Inc. (2005).
\newblock Paper 113-30

\bibitem{Allison2009}
Allison, P.D.: Missing data.
\newblock In: R.E.~Millsa, A.~Maydeu-Olivares (eds.) The SAGE Handbook of Quantitative Methods in psychology, chap.~4, pp. 73--90. SAGE Publications Ltd., London
  (2009).
\newblock \doi{10.4135/9780857020994.n4}

\bibitem{amigo2009}
Amig{\'o}, E., Gonzalo, J., Artiles, J., Verdejo, F.: A comparison of extrinsic
  clustering evaluation metrics based on formal constraints.
\newblock Inf Retrieval \textbf{12}, 461--486 (2009)

\bibitem{cora}
Cora citation matching data set.
\newblock \url{http://people.cs.umass.edu/~mccallum/data.html}.
\newblock Last accessed: 24 March 2014

\bibitem{baeza1999modern}
Baeza-Yates, R., Ribeiro-Neto, B., et~al.: Modern information retrieva.
\newblock ACM press New York (1999)

\bibitem{BanoAzam2015}
Bano, H., Azam, F.: Innovative windows for duplicate detection.
\newblock International Journal of Software Engineering and Its Applications
  \textbf{9}(1), 95--104 (2015)

\bibitem{Bilenko2003}
Bilenko, M., Mooney, R.J.: Adaptive duplicate detection using learnable string
  similarity measures.
\newblock In: Proceedings of the ninth ACM SIGKDD international conference on
  Knowledge discovery and data mining, pp. 39--48. ACM (2003)

\bibitem{bilenko2003adaptive}
Bilenko, M., Mooney, R.J.: Adaptive duplicate detection using learnable string
  similarity measures.
\newblock In: Proceedings of the Ninth ACM SIGKDD International Conference on
  Knowledge Discovery and Data Mining, pp. 39--48. ACM (2003)

\bibitem{bilenko2003evaluation}
Bilenko, M., Mooney, R.J.: On evaluation and training-set construction for
  duplicate detection.
\newblock In: Proceedings of the KDD-2003 Workshop on Data Cleaning, Record
  Linkage, and Object Consolidation, pp. 7--12 (2003)

\bibitem{BleiNgJordan2003}
Blei, D.M., Ng, A.Y., Jordan, M.I.: Latent Dirichlet allocation.
\newblock Journal of Machine Learning Research \textbf{3}, 993--1022 (2003)

\bibitem{chen2010nmi}
Chen, M.: Normalized mutual information.
\newblock
  \url{http://www.mathworks.com/matlabcentral/fileexchange/29047-normalized-mutual-information}
  (2010).
\newblock Contact: mochen@ie.cuhk.edu.hk; Last accessed: 26 February 2015

\bibitem{ChiZhu2017}
Chi, L., Zhu, X.: Hashing Techniques: A Survey and Taxonomy.
\newblock ACM Comput. Surv. \textbf{50}(1), 11:1--11:36 (2017)

\bibitem{Christen2012}
Christen, P.: A survey of indexing techniques for scalable record linkage and
  deduplication.
\newblock IEEE Transactions on Knowledge and Data Engineering \textbf{24}(9),
  1537--1555 (2012)

\bibitem{soft_tfidf_original_cohen}
Cohen, W.W., Ravikumar, P.D., Fienberg, S.E., et~al.: A comparison of string
  distance metrics for name-matching tasks.
\newblock In: IIWEB'03 Proceedings of the 2003 International Conference on Information Integration on the Web, pp. 73--78 (2003)

\bibitem{cohen2002learning}
Cohen, W.W., Richman, J.: Learning to match and cluster large high-dimensional
  data sets for data integration.
\newblock In: Proceedings of the Eighth ACM SIGKDD International Conference on
  Knowledge Discovery and Data Mining, pp. 475--480. ACM (2002)

\bibitem{DeVries2011}
De~Vries, T., Ke, H., Chawla, S., Christen, P.: Robust record linkage blocking
  using suffix arrays and bloom filters.
\newblock ACM Transactions on Knowledge Discovery from Data (TKDD)
  \textbf{5}(2), 9 (2011)

\bibitem{Draisbachetal2011}
Draisbach, U., Naumann, F.: A generalization of blocking and windowing
  algorithms for duplicate detection.
\newblock In: Data and Knowledge Engineering (ICDKE), 2011 International
  Conference on, pp. 18--24. IEEE (2011)

\bibitem{Dumais2004}
Dumais, S.T.: Latent semantic analysis.
\newblock Ann. Rev. Info. Sci. Tech. \textbf{38}(1), pp. 188--230 (2004)

\bibitem{elmagarmid2007duplicate}
Elmagarmid, A.K., Ipeirotis, P.G., Verykios, V.S.: Duplicate record detection:
  A survey.
\newblock Knowledge and Data Engineering, IEEE Transactions on \textbf{19}(1),
  1--16 (2007)

\bibitem{fiedler2010cell}
Fiedler, S.: Cell array to {CSV}-file [cell2csv.m], updated.
\newblock
  \url{http://uk.mathworks.com/matlabcentral/fileexchange/4400-cell-array-to-csv-file--cell2csv-m-}
  (2010).
\newblock Modified by Rob Kohr; last accessed: 4 September 2014

\bibitem{friedman2001elements}
Friedman, J., Hastie, T., Tibshirani, R.: The elements of statistical learning.
\newblock Springer series in statistics New York (2001)

\bibitem{fu2012multiple}
Fu, Z., Zhou, J., Christen, P., Boot, M.: Multiple instance learning for group
  record linkage.
\newblock Advances in Knowledge Discovery and Data Mining pp. 171--182 (2012)

\bibitem{vanGennipHunteretal13}
van Gennip, Y., Hunter, B., Ahn, R., Elliott, P., Luh, K., Halvorson, M., Reid,
  S., Valasik, M., Wo, J., Tita, G.E., Bertozzi, A.L., Brantingham, P.J.:
  Community detection using spectral clustering on sparse geosocial data.
\newblock SIAM J. Appl. Math. \textbf{73}(1), 67--83 (2013)

\bibitem{gonzalez2008}
Gonz{\'a}lez, E., Turmo, J.: Non-parametric document clustering by ensemble
  methods.
\newblock Procesamiento del Lenguaje Natural \textbf{40}, 91--98 (2008)

\bibitem{hall2010privacy}
Hall, R., Fienberg, S.E.: Privacy-preserving record linkage.
\newblock  In: Domingo-Ferrer J., Magkos E. (eds) Privacy in Statistical Databases. Lecture Notes in Computer Science, \textbf{6344}. Springer, Berlin, Heidelberg (2010)

\bibitem{harris1999}
Harris, M., Aubert, X., Haeb-Umbach, R., Beyerlein, P.: A study of broadcast
  news audio stream segmentation and segment clustering.
\newblock In: Proceedings of EUROSPEECH’99, pp. 1027--–1030 (1999)

\bibitem{Hassanzadeh2009}
Hassanzadeh, O., Chiang, F., Lee, H.C., Miller, R.J.: Framework for evaluating
  clustering algorithms in duplicate detection.
\newblock Proceedings of the VLDB Endowment \textbf{2}(1), 1282--1293 (2009)

\bibitem{horton2007much}
Horton, N.J., Kleinman, K.P.: Much ado about nothing.
\newblock The American Statistician \textbf{61}(1) (2007)

\bibitem{Huisman2009}
Huisman, M.: Imputation of missing network data: Some simple procedures.
\newblock Journal of Social Structure \textbf{10}(1), 1--29 (2009)

\bibitem{jaro_original}
Jaro, M.A.: Advances in record-linkage methodology as applied to matching the
  1985 census of tampa, florida.
\newblock Journal of the American Statistical Association \textbf{84}(406),
  414--420 (1989)

\bibitem{jaro95}
Jaro, M.A.: Probabilistic linkage of large public health data file.
\newblock In: Statistics in Medicine, \textbf{14}, pp. 491--498 (1995)

\bibitem{RameshKannanetal2016}
Kannan, R.R., Abarna, D., Aswini, G., Hemavathy, P.: Effective progressive
  algorithm for duplicate detection on large dataset.
\newblock International Journal of Scientific Research in Science and
  Technology \textbf{2}(2), 105--110 (2016)

\bibitem{KimGolubPark2005}
Kim, H., Golub, G.H., Park, H.: Missing value estimation for dna microarray
  gene expression data: local least squares imputation.
\newblock Bioinformatics \textbf{21}(2), 187--198 (2005).
\newblock \doi{10.1093/bioinformatics/bth499}.
\newblock \urlprefix\url{+ http://dx.doi.org/10.1093/bioinformatics/bth499}

\bibitem{KimGolubPark2006}
Kim, H., Golub, G.H., Park, H.: Missing value estimation for {DNA} microarray
  gene expression data: local least squares imputation.
\newblock Bioinformatics \textbf{22}(11), 1410--1411 (2006).
\newblock \doi{10.1093/bioinformatics/btk053}.
\newblock \urlprefix\url{+ http://dx.doi.org/10.1093/bioinformatics/btk053}

\bibitem{KimPark2008}
Kim, H., Park, H.:  Nonnegative matrix factorization based on alternating nonnegativity constrained least squares and active set method.
\newblock SIAM Journal on Matrix Analysis and Applications \textbf{30}(2), 713--730 (2008)

\bibitem{koehler2004matrix}
Koehler, M.: matrix2latex, updated.
\newblock
  \url{http://www.mathworks.com/matlabcentral/fileexchange/4894-matrix2latex}
  (2004).
\newblock Last accessed: 24 March 2014

\bibitem{komarov2010set}
Komarov, O.: Set functions with multiple inputs, updated.
\newblock
  \url{http://uk.mathworks.com/matlabcentral/fileexchange/28341-set-functions-with-multiple-inputs/content/SetMI/unionm.m}
  (2010).
\newblock Last accessed: 24 March 2014

\bibitem{larose2014discovering}
Larose, D.T., Larose, C.D.: Discovering knowledge in data: an introduction to
  data mining, 2nd edn.
\newblock John Wiley \& Sons, Inc., Hoboken, New Jersey (2014)

\bibitem{Layek2016}
Layek, A.K., Gupta, A., Ghosh, S., Mandal, S.: Fast near-duplicate detection
  from image streams on online social media during disaster events.
\newblock In: India Conference (INDICON), 2016 IEEE Annual, pp. 1--6. IEEE
  (2016)

\bibitem{Leitaoetal2013}
Leitao, L., Calado, P., Herschel, M.: Efficient and effective duplicate
  detection in hierarchical data.
\newblock IEEE Transactions on Knowledge and Data Engineering \textbf{25}(5),
  1028--1041 (2013)

\bibitem{manning2008introduction}
Manning, C.D., Raghavan, P., Sch{\"u}tze, H.: Introduction to information
  retrieval, vol.~1.
\newblock Cambridge University Press Cambridge (2008)

\bibitem{mccallum2000}
McCallum, A.K., Nigam, K., Rennie, J., Seymore, K.: Automating the construction
  of internet portals with machine learning.
\newblock Journal of Information Retrieval \textbf{3}(2) (2000)

\bibitem{meila2007}
Meil\u{a}, M.: Comparing clusterings --- an information based distance.
\newblock J. Multivariate Anal. \textbf{98}, 873--–895 (2007)

\bibitem{monge1997efficient}
Monge, A.E., Elkan, C.P.: Efficient domain-independent detection of
  approximately duplicate database records.
\newblock In: Proc. of the ACM-SIGMOD Workshop on Research Issues in on
  Knowledge Discovery and Data Mining (1997)

\bibitem{introdupdet}
Naumann, F., Herschel, M.: An introduction to duplicate detection.
\newblock Synthesis Lectures on Data Management \textbf{2}(1), 1--87 (2010)

\bibitem{nuray2013adaptive}
Nuray-Turan, R., Kalashnikov, D.V., Mehrotra, S.: Adaptive connection strength
  models for relationship-based entity resolution.
\newblock Journal of Data and Information Quality (JDIQ) \textbf{4}(2), 8
  (2013)

\bibitem{PangJinJiang2015}
Pang, G., Jin, H., Jiang, S.: CenKNN: a scalable and effective text classifier.
\newblock Data Mining and Knowledge Discovery \textbf{29}(3), 593--625 (2015)

\bibitem{Papadakis2013}
Papadakis, G., Ioannou, E., Palpanas, T., Niederee, C., Nejdl, W.: A blocking
  framework for entity resolution in highly heterogeneous information spaces.
\newblock IEEE Transactions on Knowledge and Data Engineering \textbf{25}(12),
  2665--2682 (2013)

\bibitem{Papadakis2011}
Papadakis, G., Nejdl, W.: Efficient entity resolution methods for heterogeneous
  information spaces.
\newblock In: Data Engineering Workshops (ICDEW), 2011 IEEE 27th International
  Conference on, pp. 304--307. IEEE (2011)

\bibitem{Papenbrock2015}
Papenbrock, T., Heise, A., Naumann, F.: Progressive duplicate detection.
\newblock IEEE Transactions on Knowledge and Data Engineering \textbf{27}(5),
  1316--1329 (2015)

\bibitem{pigott2001review}
Pigott, T.D.: A review of methods for missing data.
\newblock Educational Research and Evaluation \textbf{7}(4), 353--383 (2001)

\bibitem{Ramya2017}
Ramya, R., Venugopal, K., Iyengar, S., Patnaik, L.: Feature extraction and
  duplicate detection for text mining: A survey.
\newblock Global Journal of Computer Science and Technology \textbf{16}(5)
  (2017)

\bibitem{datasets}
Duplicate detection, record linkage, and identity uncertainty: Datasets.
\newblock \url{http://www.cs.utexas.edu/users/ml/riddle/data.html}.
\newblock Last accessed: 24 March 2014

\bibitem{vanrijsbergen1979}
van Rijsbergen, C.J.: Information Retrieval, 2nd edn.
\newblock Butterworth-Heinemann, Newton, MA, USA (1979)

\bibitem{salton1988term}
Salton, G., Buckley, C.: Term-weighting approaches in automatic text retrieval.
\newblock Information Processing \& Management \textbf{24}(5), 513--523 (1988)

\bibitem{vsm_original}
Salton, G., Wong, A., Yang, C.S.: A vector space model for automatic indexing.
\newblock Commun. ACM \textbf{18}(11), 613--620 (1975)

\bibitem{scannapieco2007privacy}
Scannapieco, M., Figotin, I., Bertino, E., Elmagarmid, A.K.: Privacy preserving
  schema and data matching.
\newblock In: Proceedings of the 2007 ACM SIGMOD International Conference on
  Management of Data, pp. 653--664. ACM (2007)

\bibitem{strehl2002}
Strehl, A., Ghosh, J.: Cluster ensembles --- a knowledge reuse framework for
  combining multiple partitions.
\newblock Journal of Machine Learning Research \textbf{3}, 583--617 (2002)

\bibitem{Tamilselvi2011}
Tamilselvi, J.J., Gifta, C.B.: Handling duplicate data in data warehouse for
  data mining.
\newblock International Journal of Computer Applications (0975--8887)
  \textbf{15}(4), 1--9 (2011)

\bibitem{tejada01}
Tejada, S., Knoblock, C.A., Minton, S.: Learning object identification rules
  for information integration.
\newblock Information Systems \textbf{26}(8), 607--633 (2001)

\bibitem{traud2011comparing}
Traud, A.L., Kelsic, E.D., Mucha, P.J., Porter, M.A.: Comparing community
  structure to characteristics in online collegiate social networks.
\newblock SIAM Review \textbf{53}(3), 526--543 (2011)

\bibitem{traud2011zrand}
Traud, A.L., Kelsic, E.D., Mucha, P.J., Porter, M.A.: zrand.
\newblock \url{http://netwiki.amath.unc.edu/GenLouvain/GenLouvain} (2011).
\newblock Last accessed: 24 March 2014

\bibitem{tromp2006record}
Tromp, M., Reitsma, J., Ravelli, A., M{\'e}ray, N., Bonsel, G.: Record linkage:
  making the most out of errors in linking variables.
\newblock In: AMIA Annual Symposium Proceedings, \textbf{2006}, p. 779. American
  Medical Informatics Association (2006)

\bibitem{Watadaetal2016}
Watada, J., Shi, C., Yabuuchi, Y., Yusof, R., Sahri, Z.: A rough set approach
  to data imputation and its application to a dissolved gas analysis dataset.
\newblock In: Computing Measurement Control and Sensor Network (CMCSN), 2016
  Third International Conference on, pp. 24--27. IEEE (2016)

\bibitem{Whang2013}
Whang, S.E., Marmaros, D., Garcia-Molina, H.: Pay-as-you-go entity resolution.
\newblock IEEE Transactions on Knowledge and Data Engineering \textbf{25}(5),
  1111--1124 (2013)

\bibitem{winkler_original}
Winkler, W.: String comparator metrics and enhanced decision rules in the
  fellegi-sunter model of record linkage.
\newblock In: Proceedings of the Section on Survey Research Methods, pp.
  354--359. (American Statistical Association) (1990)

\bibitem{winkler1999state}
Winkler, W.E.: The state of record linkage and current research problems.
\newblock In: Statistical Research Division, US Census Bureau (1999)

\bibitem{Winkler02methodsfor}
Winkler, W.E.: Methods for record linkage and {B}ayesian networks.
\newblock Tech. rep., Series RRS2002/05, U.S. Bureau of the Census (2002)

\bibitem{Winkler06overviewof}
Winkler, W.E.: Overview of record linkage and current research directions.
\newblock Tech. rep., Bureau of the Census (2006)

\bibitem{Xiao2011}
Xiao, C., Wang, W., Lin, X., Yu, J.X., Wang, G.: Efficient similarity joins for
  near-duplicate detection.
\newblock ACM Transactions on Database Systems (TODS) \textbf{36}(3), 15 (2011)

\end{thebibliography}

\end{document}